\documentclass[12pt,a4paper]{article}

\usepackage{graphicx}
\usepackage{xcolor}
\usepackage{geometry}
\usepackage{amsmath}
\usepackage{amssymb}
\usepackage{booktabs}
\usepackage{longtable}
\usepackage{array}
\usepackage{enumitem}
\usepackage{setspace}
\usepackage{microtype}
\usepackage{fancyhdr}
\usepackage{titlesec}
\usepackage{multirow}
\usepackage{mathtools}
\usepackage{bm}
\usepackage{float}
\usepackage[format=hang, font={small, it}, labelfont=bf]{caption}
\usepackage{algorithm}
\usepackage{algorithmic}
\usepackage{makecell}
\usepackage{ragged2e}    
\usepackage{array}         
\usepackage{booktabs}

\usepackage{mathrsfs}
\usepackage{xfrac}
\usepackage[overload]{empheq}
\usepackage{authblk}
\usepackage{tabularx}

\usepackage[square,numbers, sort&compress]{natbib}
\titleformat{\section}{\large\bfseries}{\thesection}{1em}{}
\titleformat{\subsection}{\normalsize\bfseries}{\thesubsection}{1em}{}
\titleformat{\subsubsection}{\normalsize\itshape}{\thesubsubsection}{1em}{}

\usepackage{hyperref}
\hypersetup{colorlinks=true, linkcolor=blue, citecolor=blue, urlcolor=blue}

\numberwithin{equation}{section}
\allowdisplaybreaks[4]

\begin{document}

\begin{titlepage}
\centering
\vspace*{2.5cm}

{\Large\bfseries{When Certainty Emerges from Stochasticity: Hidden Attractor of Deterministic Motion}\\[1.4cm]}
\vspace{2cm}

{
\large
\textbf{D.Y. ZHONG}$^{*1,2}$
}

\vspace{.5cm}

{
\normalsize
$^{1}$State Key Laboratory of Hydroscience and Engineering, Tsinghua University, Beijing 100084, China\\
$^{2}$Department of Hydraulic Engineering, Tsinghua University, Beijing 100084, China
}

\vspace{0.5cm}

{
\normalsize 
\textit{$^{*}:$ Corresponding author: zhongdy@tsinghua.edu.cn}
}

\vfill

{
\large 
Jun $19^{th}$ 2026
}

\end{titlepage}

\newpage
\section*{Abstract}

Macroscopic deterministic motion is traditionally interpreted as a result of statistical averaging. In this paper, we show that it is a strict geometric attractor of the contact flow. We reveal a contact constraint mechanism where the exponential amplification of probability gradients is exactly counterbalanced by the decay of second-order contact stiffness, forcing the macroscopic-microscopic coupling to vanish. This coupling acts as a Jacobi field, which decays in dissipative systems to enable deterministic focusing. We construct the contact potential via an invariant-measure construction, unifying the treatment of point attractors, limit cycles, and chaotic systems. Unlike the Mori-Zwanzig projection, this approach strictly conserves information, showing that determinism arises from the geometric reorganisation of information rather than its loss.

\addcontentsline{toc}{section}{Abstract}

\vspace{0.5cm}

\noindent\textbf{Keywords}:

\noindent contact potential; contact constraint; deterministic focusing; Jacobi field; dissipative systems

\newpage
\tableofcontents
\listoftables

\newpage

\section{Introduction}\label{sec:introduction}

Macroscopic deterministic laws govern the evolution of the macro-world and are among the most precisely tested in physics. Despite persistent microscopic perturbations and thermal noise, macroscopic deterministic equations maintain their structural stability with remarkable resilience and thus make the real world understandable \cite{van1992stochastic}. 

In classical statistical mechanics, this stability is understood as a mathematical result: the irregular motion of vast numbers of microscopic constituents, when subjected to some sort of averaging, yields smooth macroscopic behaviour through the suppression of fluctuations \cite{zwanzig2001nonequilibrium}. This statistical principle, that determinism arises as a limit of stochasticity via the law of large numbers, has served as the operational foundation of non-equilibrium statistical physics for hundreds of years, achieving remarkable success in equilibrium settings and in systems possessing a clear separation of time scales \cite{Pope_2000, ZHONG-2022-KINETICEQAUTION, Zhong_2024}.

While the law of large numbers provides the mathematical basis for the suppression of fluctuations, it does not itself constitute a physical mechanism. It states that fluctuations vanish in the limit of large ensembles, leaving deterministic macroscopic equations behind, but it does not explain why those particular equations emerge, nor does it account for why deterministic attractors in dissipative systems maintain their structure against continuous microscopic perturbation. For example, the Mori-Zwanzig (MZ) formalism, the standard operator-theoretic approach to coarse-graining \cite{Mori1965, Zwanzig1960, Zwanzig1961}, recovers deterministic equations by projecting out irrelevant variables and approximating the memory kernel. The modeller must prescribe the relevant variables in advance; the framework offers no intrinsic criterion for selecting the drift field from the underlying stochastic dynamics. Consequently, within this framework, determinism is effectively assumed rather than derived.

The limitations of this statistical picture are most visible in far-from-equilibrium dissipative systems, where deterministic attractors form and persist under noise across a wide range of phenomena including self-organising structures, coherent turbulence, and chaotic dynamics \cite{Cross1993, nicolis1977Self-organization, Perc2006From, Erdmann2000Brownian}. Such robustness suggests that determinism is not merely a statistical residue of ensemble averaging, but a structural feature of the stochastic dynamics itself. It indicates the presence of an underlying mechanism that isolates microscopic stochasticity from macroscopic dynamics, a structure not captured by traditional statistical formulations because the natural geometric language, symplectic geometry, cannot accommodate it structurally.

Symplectic geometry provides the rigorous mathematical foundation for classical Hamiltonian mechanics, but is structurally constrained to conservative, deterministic flows \cite{Abraham1978}. It cannot accommodate dissipative effects, irreversible evolution, or the probabilistic degrees of freedom that characterise stochastic systems. This limitation is not technical but fundamental: the symplectic form is closed, encoding energy conservation and time-reversibility, whereas stochastic dissipative dynamics is inherently non-conservative and irreversible. The contact geometry of stochastic vector bundles, introduced in our previous construction \cite{ZHONG2025}, provides a natural geometric language for formulating this underlying structure.

Contact geometry, the odd-dimensional counterpart to symplectic geometry, is distinguished by a canonical contact form that is intrinsically non-closed \cite{anold2010, ArnoldVI1989, Bravetti2017}. This non-closure provides the natural geometric encoding of dissipation and irreversibility. Within the contact manifold constructed over the stochastic vector bundle, the system's physical state, its stochastic fluctuations, dissipative effects, and changes in probability are unified into a single geometric object governed by the least constraint theorem \cite{ZHONG2025}, the dissipative stochastic counterpart to Hamilton's principle. 

Building on the contact-geometric formulation of stochastic dynamics \cite{ZHONG2025}, this work reveals that macroscopic deterministic motion is not a simple result of statistical averaging, but a strict geometric attractor. Our central discovery is a hidden, exact balancing mechanism: the exponential amplification of microscopic probability fluctuations is perfectly counterbalanced by the decay of the system’s internal stiffness, forcing their effective interaction to vanish exponentially. This finding fundamentally shifts our understanding of certainty: determinism emerges from the strict conservation and geometric reorganisation of information. To illustrate this mechanism, we construct a unified geometric potential from the system’s invariant measure, which seamlessly applies to point attractors, limit cycles, and chaotic systems. Furthermore, this deterministic separation naturally arises in dissipative systems where energy dispersal breaks the microscopic-macroscopic coupling. 

Most importantly, this study shifts the origin of determinism from information loss, as in the Mori--Zwanzig projection, to information conservation under the contact constraint. The contact-geometric structure offers a fundamentally different approach compared to the Mori--Zwanzig projection formalism \cite{Mori1965, Zwanzig1960, Zwanzig1961}, the standard operator-theoretic approach to coarse-graining. In the Mori--Zwanzig framework, determinism is recovered by discarding information, projecting out irrelevant variables and approximating the memory kernel. In the contact-geometric framework, no information is discarded: the constraint function is strictly conserved, and deterministic focusing emerges from the geometric reorganisation of information under the contact constraint, rather than from its loss. This distinction, which shifts from information loss to information conservation and from statistical approximation to a geometric attractor, forms the conceptual basis of this work.

The remainder of the paper is organised as follows. Section~\ref{sec:Fundation} introduces the geometric language, contact dynamics, in which the structure becomes visible. Section~\ref{sec:Exact-Closure} establishes that this language possesses an exact closure, ensuring that what is revealed is not an artefact of approximation. Section~\ref{sec:Emergence-Determinism} presents the structure itself: the contact constraint, the Jacobi field, and the deterministic focusing mechanism. Section~\ref{sec:examples} verifies the mechanism on four representative systems: a linear dissipative system, the van der Pol oscillator, the Lorenz system, and the harmonic oscillator, one for each dynamical regime. Section~\ref{sec:Discussion} places this structure in context against the Mori--Zwanzig framework. Section~\ref{sec:conclusions} summarises the implications.

\section{Foundational Framework of Contact Dynamics}\label{sec:Fundation}
This section introduces the contact manifold, the natural geometric setting in which deterministic motion emerges not as an approximation but as a strict attractor. The concepts and equations are based on \cite{ZHONG2025}; we restate only the essential elements to keep the presentation self-contained. Herein and hereafter, the Einstein summation convention is adopted throughout the paper, where repeated indices imply summation over the full space.

\subsection{Basic Concepts of Contact Manifold and Dynamics}

\subsubsection{Stochastic vector bundles}
Let $(\Omega, \mathcal{F}, \mathbb{P})$ denote a complete filtered probability space, where $\Omega$ is the sample space, $\mathcal{F}$ is the $\sigma$-algebra of measurable events, $\{\mathcal{F}_t\}_{t \in T} \subseteq \mathcal{F}$ is the filtration adapted to the stochastic evolution, and $\mathbb{P}$ is the space of probability measure. $M$ denotes an $m$-dimensional smooth base manifold with local coordinates $x^i \in \mathbb{R}^m$ ($i=1,2,\dots,m$).

A stochastic process is an $\mathcal{F}_t$-measurable map $\gamma: \Omega \hookrightarrow M$ that maps sample outcomes to points on the base manifold. A stochastic vector bundle is a smooth vector bundle $\pi: E \to M$, where the total space $E$ is the collection of all realizations (sample paths) of the stochastic process, equipped with a measurable section (stochastic vector field) $Y: U\subset M \to E$ defined on an open subset $U$ satisfying the bundle projection condition $\pi \circ Y = \mathrm{Id}_U$.

The local section $Y$ is assigned a probability measure $P$ on the total space $E$: for an open set $U_y \subset E$ around $y$, the probability of the stochastic variable taking values in $U_y$ is
\begin{equation}
    P(Y \in U_y) \equiv {P}\left(\gamma^{-1}(Y^{-1}(U_y))\right).
    \label{eq:induced_prob_measure}
\end{equation}
For simplicity, it is denoted as $P(y)$ in the following study.

The probability measure space $\mathbb{P}$ of the stochastic vector bundle admits a natural infinite-order jet bundle structure $J^\infty(E, \mathbb{P})$, of which the local coordinates are given by $(y^i, P, P_{\mu_1}, P_{\mu_1\mu_2}, \dots, P_{\mu_1\mu_2\cdots\mu_\infty})$, where $P_{\mu_1\cdots\mu_k} = \partial_{\mu_1}\cdots\partial_{\mu_k} P(y)$ denotes the $k$-th order partial derivative of the probability measure with respect to the base coordinates.

\subsubsection{Contact manifold and coordinate setting}
Based on the stochastic vector bundle $\pi: E \to M$ and its infinite-order jet bundle $J^\infty(E, \mathbb{P})$, we define the $(2n+1)$-dimensional contact manifold $(T^*E\times \mathbb{R}, \Theta)$ reduced from $J^\infty(E, \mathbb{P})$ via contact mapping, with the following coordinate system and core geometric properties: 
\begin{enumerate}[label=(\roman*)]
\item base coordinates $y^i \in \mathbb{R}^n$ ($i=1,2,\dots,n$), which are the state variables of the stochastic system inherited from the base manifold $M$; 
\item fibre coordinates $\phi_i \in \mathbb{R}^n$ ($i=1,2,\dots,n$), which are the conjugate variables of the base coordinates encoding the local connection of the stochastic vector bundle, capturing the stochastic fluctuation, dissipation and path dependence of the system, given by $\phi=\phi_i dy^i= \sum_{1\le k \le \infty}\sum_{\mu_1<\cdots<\mu_k}B_i^{\mu_1\cdots\mu_k}P_{\mu_1\cdots\mu_k}dy^i$; and 
\item the evolution parameter $t \in \mathbb{R}$, which is the time variable governing the stochastic dynamics.
\end{enumerate}

To ensure dimensional consistency, all quantities in the subsequent analysis are rendered dimensionless by characteristic scales $Y_c$ for the base coordinates, $T_c$ for time, and $P_c$ for probability density, among others. For instance, $\tilde{y}=y/Y_c$ and $\tilde{t}=t/T_c$. For brevity, the tildes are omitted and all variables are treated as dimensionless.

The contact structure of the manifold is characterised by the non-closed, completely non-integrable canonical contact 1-form, which has the equivalent form
\begin{equation}
    \Theta =  H(t,y,\phi) dt - \phi_i dy^i,
    \label{eq:contact_1form_time}
\end{equation}
where the contact probability potential $H(t,y,\phi) = dP(\dot{y}) = \dot{y}^i \frac{\partial P}{\partial y^i}$ physically represents the rate of change of the probability measure along the system's evolution trajectory, encoding the information exchange between the stochastic system and its environment.

The contact 1-form $\Theta$ satisfies the following conditions to form a well-defined contact structure: 
\begin{enumerate}[label=(\roman*)]
\item Non-degeneracy: $\Theta \neq 0$ everywhere on the manifold; 
\item Volume form condition: $\Theta \wedge (d\Theta)^n \neq 0$, i.e., the top-form is a nowhere-vanishing volume form on the $(2n+1)$-dimensional manifold; 
\item Complete non-integrability: the contact distribution $\ker(\Theta)$ satisfies $[\ker(\Theta), \ker(\Theta)] \not\subset \ker(\Theta)$, which geometrically encodes the path dependence and irreversibility of stochastic dynamics.
\end{enumerate}

The contact 1-form $\Theta$ induces a canonical Whitney sum decomposition of the tangent space of the extended phase space $\mathcal{E}=T^*E\otimes \mathbb{R}$ at each point of the manifold:
\begin{equation}
    T\mathcal{E} = \ker(\Theta) \oplus \ker(d\Theta),
    \label{eq:tangent_space_decomposition}
\end{equation}
where $\ker(\Theta)$ is the contact distribution capturing the probability evolution of the system, and $\ker(d\Theta)$ is the subspace of vector fields preserving the contact structure.

\subsubsection{Contact dynamics}
For the contact manifold $(\mathcal{E}, \Theta)$ with smooth contact probability potential $H(t,y,\phi)$, the stochastic evolution of the system is governed by the contact dynamical equations, which are derived from the unique contact vector field preserving the contact structure.

The standard form of the contact dynamical equations is
\begin{equation}
    \begin{cases}
        \displaystyle \dot{y}^i =\;\;\, \frac{\partial H}{\partial \phi_i}, \\[6pt]
        \displaystyle \dot{\phi}_i = -\frac{\partial H}{\partial y^i}, \\[6pt]
        \displaystyle \frac{dH}{dt} = \frac{\partial H}{\partial t},
    \end{cases}
    \label{eq:contact_hamiltonian_standard}
\end{equation}
where the first two equations describe the co-evolution of the system state and the local connection, and the third equation characterises the time dependence of the contact probability potential.

Defining the canonical Poisson bracket for smooth functions $F, G$ on the contact manifold as
\begin{equation}
    \{F, G\} \equiv \frac{\partial F}{\partial y^i}\frac{\partial G}{\partial \phi_i} - \frac{\partial F}{\partial \phi_i}\frac{\partial G}{\partial y^i},
    \label{eq:poisson_bracket_def}
\end{equation}
the contact dynamical equations can be rewritten in the compact Poisson bracket form:
\begin{equation}
    \begin{cases}
        \displaystyle \frac{dy^i}{dt} = \{y^i, H\}, \\[6pt]
        \displaystyle \frac{d\phi_i}{dt} = \{\phi_i, H\}, \\[6pt]
        \displaystyle \frac{\partial \varepsilon}{\partial t} = -\{\varepsilon, H\},
    \end{cases}
    \label{eq:contact_hamiltonian_poisson}
\end{equation}
where $\varepsilon$ is the constraint function defined below in equation~\eqref{eq:constraint_function_def}.

The flow generated by the contact dynamical equations satisfies the Lie derivative condition $\mathcal{L}_{X_H} \Theta = 0$, meaning the contact structure of the manifold is strictly preserved along the system's evolution, which is the geometric counterpart of Liouville's theorem in symplectic mechanics for dissipative stochastic systems.

\subsubsection{Constraint function}
For the contact manifold $(\mathcal{E}, \Theta)$ of the stochastic vector bundle, the constraint function $\varepsilon \in C^\infty(\mathcal{E})$ is a smooth scalar function defined as:
\begin{equation}
    \varepsilon = \phi_i \dot{y}^i - dP(\dot{y}) = \phi_i \dot{y}^i - H,
    \label{eq:constraint_function_def}
\end{equation}
with the following core geometric and physical properties. It quantifies the deviation of the system's evolution trajectory from the contact distribution $\ker(\Theta)$, and geometrically encodes both the dissipation and stochastic noise of the system. It is invariant along the flow of the contact vector field, i.e., $$\frac{d\varepsilon}{dt} = \mathcal{L}_{X_H} \varepsilon = 0$$, serving as a first integral of the open stochastic system. It is not an auxiliary constraint but generates intrinsic geometric constraint forces that govern the system's dynamics, arising from the contact structure and the stochastic nature of the system, thus unifying the effects of dissipation, noise, and geometric constraints into a single deterministic force term.

\subsubsection{Least constraint theorem for stochastic vector bundles}

For the contact manifold $(\mathcal{E}, \Theta)$ and the smooth constraint function $\varepsilon$, there exists a unique smooth vector field $X_H \in \ker(d\Theta) \subset T\mathcal{E}$ satisfying:
\begin{equation}
    \iota_{X_H} \Theta = -\varepsilon, \quad \iota_{X_H} d\Theta = 0.
    \label{eq:contact_vector_field_condition}
\end{equation}
The flow generated by this vector field strictly preserves the contact structure of the manifold and uniquely determines the evolution trajectory of the stochastic system.

The flow generated by the contact vector field $X_H$ corresponds to the extremal paths of the constraint action functional, which satisfies the variational principle:
\begin{equation}
    \delta S = \delta \int_{t_1}^{t_2} \varepsilon \, dt = 0.
    \label{eq:least_constraint_variational}
\end{equation}
This principle is the dissipative and stochastic counterpart of the least action principle in classical mechanics. It states that the stochastic system evolves along trajectories that extremise the time integral of the constraint function, which optimises the trade-off between energy dissipation and probabilistic fluctuations during state transitions.

The least constraint theorem unifies the underdamped, overdamped and critically damped regimes of stochastic dissipative systems through a single constraint function $\varepsilon$. The contact dynamical equations \eqref{eq:contact_hamiltonian_standard} and \eqref{eq:contact_hamiltonian_poisson} can be directly derived from the variational principle \eqref{eq:least_constraint_variational}, establishing a consistent geometric framework for the analysis of non-equilibrium stochastic systems.

\subsection{Constraint Function and Master Equation}

The Euler operator is defined as $E := \phi_i \frac{\partial}{\partial \phi_i}$, and the operator $D$ is defined as $D := E - 1$. By the Euler operator, the constraint function $\varepsilon$ defined in equation~\eqref{eq:constraint_function_def} can be equivalently expressed as the action of the operator $D$ on the contact potential:
\begin{equation}
\varepsilon = D[H] = \phi_i \frac{\partial H}{\partial \phi_i} - H.
\label{eq:constraint_function}
\end{equation}

The total time derivative of the constraint function along the contact Hamiltonian flow satisfies
\begin{equation}
\frac{d\varepsilon}{dt} = \frac{\partial \varepsilon}{\partial t} + \{\varepsilon, H\},
\label{eq:epsilon_evolution}
\end{equation}
where the canonical Poisson bracket $\{\cdot,\cdot\}$ is as defined in \eqref{eq:poisson_bracket_def}.
The conservation condition $d\varepsilon/dt=0$ is equivalent to the master equation that the contact potential must satisfy:
\begin{equation}
D\left[\frac{\partial H}{\partial t}\right] + \{D[H], H\} = 0.
\label{eq:master_equation}
\end{equation}
This equivalence follows by substituting $\varepsilon = D[H]$ and noting that $D[\partial_t H] = \phi_i \partial^2 H/\partial\phi_i\partial t - \partial H/\partial t = \partial\varepsilon/\partial t$.

\subsection{Taylor Expansion and Order-Based Spectral Separation}

The contact potential is expanded into a homogeneous Taylor series with respect to $\phi$:
\begin{equation}
H(t,y,\phi) = \sum_{n=0}^{\infty} H^{(n)}(t,y,\phi), \quad H^{(n)} = \frac{1}{n!} H^{(n) i_1 i_2 \dots i_n}(t,y) \phi_{i_1}\phi_{i_2}\dots\phi_{i_n},
\label{eq:taylor_expansion_H}
\end{equation}
where $H^{(n)}$ is a homogeneous polynomial of degree $n$ in $\phi$, and $H^{(n) i_1 \dots i_n}$ is a fully symmetric $n$-th order tensor.

By Euler's homogeneous function theorem, $E[H^{(n)}] = n H^{(n)}$, thus $D[H^{(n)}] = (n-1) H^{(n)}$. The constraint function, therefore, has the Taylor expansion
\begin{equation}
\varepsilon = D[H] = \sum_{n=0}^{\infty} (n-1) H^{(n)} = -H^{(0)} + \sum_{n=2}^{\infty} (n-1) H^{(n)}.
\label{eq:taylor_expansion_epsilon}
\end{equation}
Substituting this expansion into the master equation~\eqref{eq:master_equation} and collecting terms by the homogeneous order $p$ of $\phi$ (noting that $\{H^{(n)}, H^{(m)}\}$ is homogeneous of degree $n+m-1$), we obtain the order-by-order consistency equation for each $p=0,1,2,\dots$
\begin{equation}
(p-1) \frac{\partial H^{(p)}}{\partial t} + \sum_{n+m-1=p} (n-1) \{H^{(n)}, H^{(m)}\} = 0.
\label{eq:order_by_order_consistency}
\end{equation}

We now present the explicit form of the consistency equations for low orders, which form the basis of the subsequent construction theory. For order $p=0$, the equation reduces to the zero-order transport equation
\begin{equation}
\frac{\partial H^{(0)}}{\partial t} + H^{(1)i}(t,y) \frac{\partial H^{(0)}}{\partial y^i} = 0.
\label{eq:zero_order_transport}
\end{equation}

For order $p=1$, the equation reduces to the degeneracy condition
\begin{equation}
H^{(2) ij}(t,y) \frac{\partial H^{(0)}}{\partial y^j} = 0, \quad \forall i=1,2,\dots,m.
\label{eq:first_order_degeneracy}
\end{equation}
For order $p=2$, the equation reads
\begin{equation}
\frac{\partial H^{(2)}}{\partial t} + \{H^{(2)}, H^{(1)}\} - 3\{H^{(0)}, H^{(3)}\} = 0.
\label{eq:second_order_consistency}
\end{equation}
For $N=2$ truncation, defined as $H^{(n)}=0$ for all $n>2$, this reduces to the Lie transport equation
\begin{equation}
\frac{\partial H^{(2)}}{\partial t} + \{H^{(2)}, H^{(1)}\} = 0.
\label{eq:second_order_lie_transport}
\end{equation}
For orders $p \geq 3$, the general recurrence equation is
\begin{equation}
\begin{split}
(p-1) \frac{\partial H^{(p)}}{\partial t} &+ (p-1) \{H^{(p)}, H^{(1)}\}\\
&+ \sum_{n=2}^{p-1} (n-1) \{H^{(n)}, H^{(p+1-n)}\} \\
&- (p+1) \{H^{(0)}, H^{(p+1)}\} = 0.
\end{split}
    \label{eq:general_recurrence}
\end{equation}

\section{Construction of Contact Potentials}\label{sec:Exact-Closure}

\subsection{The First-order Degeneracy Condition}

The first-order degeneracy condition $H^{(2)}\nabla H^{(0)} = 0$ (Eq.~\eqref{eq:first_order_degeneracy}) carries profound geometric and physical implications, acting as the geometric demarcation between systems with non-constant and constant zero-order contact potential within the contact framework.

From a linear algebraic perspective, the condition implies that the gradient vector $\nabla H^{(0)}$ lies strictly within the kernel (null space) of the symmetric matrix $H^{(2)}$:
\[ \nabla H^{(0)} \in \ker(H^{(2)}) \quad \iff \quad H^{(2)} \nabla H^{(0)} = 0. \]
When $H^{(0)}$ is non-constant, $\nabla H^{(0)} \neq 0$, which forces $H^{(2)}$ to be singular. Consequently, the second-order contact stiffness tensor $H^{(2)}$ exhibits structural degeneracy, with a rank at most $m-1$.

This mathematical singularity corresponds to two different cases. First, when $\nabla H^{(0)} \neq 0$, the gradient $\nabla H^{(0)}$ is the normal vector to the level sets of $H^{(0)}$. The condition $H^{(2)} \nabla H^{(0)} = 0$ signifies that the second-order contact stiffness vanishes in the normal direction. Physically, this means the normal direction is unconstrained; the probability field possesses no restoring stiffness to contract or localise across the level surface of $H^{(0)}$. Because the governing equation (stiffness) reduces to zero, this direction becomes uncontrollable, representing a loss of a controllable degree of freedom. 

Second, when $\nabla H^{(0)} = 0$, we have $H^{(0)} = c$ (constant), so $\nabla H^{(0)} = 0$. The condition reduces to the trivial identity $H^{(2)} \cdot 0 = 0$. The matrix $H^{(2)}$ is generically non-singular (full rank). Physically, the normal restriction is entirely removed; all directions become controllable, possessing non-zero stiffness. This removal of the restriction allows the probability field to fully contract in all directions.

\subsection{Projected Lie Transport for Degeneracy Preservation}
\label{subsec:projection}

When the zero-order term $H^{(0)}$ is non-constant, the first-order degeneracy condition \eqref{eq:first_order_degeneracy} imposes that $\nabla H^{(0)}$ lies in the kernel of $H^{(2)}$, i.e.\ $H^{(2)}\nabla H^{(0)} = 0$. Geometrically, this means that the second-order contact stiffness $H^{(2)}$ is confined to the tangent space of the level sets of $H^{(0)}$. The standard Lie transport equation \eqref{eq:second_order_lie_transport}, however, treats all directions of phase space on an equal footing and does not automatically respect this tangential confinement. 

The aim of this subsection is to derive a modified transport equation that projects the dynamics onto the level sets of $H^{(0)}$, which preserves the degeneracy constraint by construction. The construction unifies the constant and non-constant cases: when $H^{(0)}=c$, the projection becomes trivial, and the modified equation reduces identically to the standard Lie transport equation.

We first recall how the standard Lie transport equation~\eqref{eq:second_order_lie_transport} emerges from the $N=2$ consistency condition, since the projected equation will be obtained by reshaping this same equation rather than by postulating a new one. For $N=2$ truncation, the order-$p=2$ consistency equation \eqref{eq:second_order_lie_transport} reads:
\begin{equation}
\frac{\partial H^{(2)}}{\partial t} + \{H^{(2)}, H^{(1)}\} = 0.
\end{equation}
Substituting $H^{(1)} = H^{(1)k} \phi_k$ into the Poisson bracket $$\{H^{(2)}, H^{(1)}\} = \frac{\partial H^{(2)}}{\partial y^k} \frac{\partial H^{(1)}}{\partial \phi_k} - \frac{\partial H^{(2)}}{\partial \phi_k} \frac{\partial H^{(1)}}{\partial y^k}$$ we obtain:
\begin{equation}
\{H^{(2)}, H^{(1)}\} 
= 
H^{(1)k} \frac{\partial H^{(2)}}{\partial y^k} 
- 
H^{(2)kl}  \phi_l M_k^m \phi_m,
\end{equation}
where $M^k_l = \partial H^{(1)k} / \partial y^l$ is the Jacobian matrix $M$. Within the contact geometric framework, $M$ governs the linearised evolution of the base manifold relative to the deterministic reference trajectory. As demonstrated by the order-by-order consistency of the master equation~\eqref{eq:master_equation}, $M$ directly dictates how probability gradients and constraint functions are transported and deformed across the spatial manifold, serving as the core operator that couples local deterministic flow to global probability topology.

Translating the second term into matrix notation, $H^{(2) kj} \phi_j M^l_k \phi_l$ represents the quadratic form $(H^{(2)} \phi)^T (M^T \phi) = \phi^T H^{(2)} M^T \phi$. Symmetrising this quadratic form yields:
\begin{equation}
\phi^T H^{(2)} M^T \phi = \frac{1}{2} \phi^T \left( M H^{(2)} + H^{(2)} M^T \right) \phi.
\end{equation}
Thus, the Poisson bracket becomes:
\begin{equation}
\{H^{(2)}, H^{(1)}\} = H^{(1)k} \frac{\partial H^{(2)}}{\partial y^k} - \frac{1}{2} \phi^T \left( M H^{(2)} + H^{(2)} M^T \right) \phi.
\end{equation}
Substituting this back into the consistency equation and defining the total derivative $\frac{DH^{(2)}}{Dt}:= \frac{\partial H^{(2)}}{\partial t} + H^{(1)k} \frac{\partial H^{(2)}}{\partial y^k}$, we match the symmetric matrices on both sides to obtain the standard tensor Lie transport equation:
\begin{equation}
\frac{DH^{(2)}}{Dt} = MH^{(2)} + H^{(2)}M^T. \label{eq:standard_lie_alt}
\end{equation}
This equation is valid regardless of whether $H^{(0)}$ is constant or non-constant. However, when $H^{(0)}$ is non-constant, it does not automatically preserve the degeneracy constraint \eqref{eq:first_order_degeneracy}, i.e., $H^{(2)}\nabla H^{(0)} = 0$. The remedy is to decompose the dynamics on the right-hand side of \eqref{eq:standard_lie_alt} into its tangential and normal components relative to the level sets of $H^{(0)}$.

Let $H^{(0)}(t,y)$ be any smooth solution of the zero-order transport equation \eqref{eq:zero_order_transport}. The natural object for separating tangential and normal directions is the orthogonal projector onto the tangent space of the level sets of $H^{(0)}$:
\begin{equation}
P_{\parallel}(t,y) :=
\begin{cases}
\displaystyle I - \frac{\nabla H^{(0)} \otimes \nabla H^{(0)}}{|\nabla H^{(0)}|^2}, & \nabla H^{(0)} \neq 0, \\[8pt]
I, & \nabla H^{(0)} = 0.
\end{cases}
\label{eq:projector}
\end{equation}
The projector $P_{\parallel}$ maps any vector in $\mathbb{R}^m$ onto the orthogonal complement $(\nabla H^{(0)})^{\parallel}$ and satisfies $P_{\parallel} \nabla H^{(0)} = 0$. When $H^{(0)}=c$ is constant, $P_{\parallel}=I$ trivially. Given the Jacobian matrix $M^i_j = \partial H^{(1)i} / \partial y^j$, the tangential part of the dynamics is captured by the projected Jacobian
\begin{equation}
\widetilde{M}(t,y) := P_{\parallel}(t,y)\, M(t,y)\, P_{\parallel}(t,y).
\label{eq:projected_jacobian}
\end{equation}

Using the orthogonal decomposition $I = P_{\parallel} + (I - P_{\parallel})$, we split the Jacobian $M$ into four blocks:
\begin{equation}
\begin{split}
M &= \bigl(P_{\parallel} + (I - P_{\parallel})\bigr) M \bigl(P_{\parallel} + (I - P_{\parallel})\bigr)\\ & = \widetilde{M} + P_{\parallel} M (I - P_{\parallel}) + (I - P_{\parallel}) M P_{\parallel} + (I - P_{\parallel}) M (I - P_{\parallel}),
\end{split} \label{eq:jacobian_decomposition}
\end{equation}
where $\widetilde{M} = P_{\parallel} M P_{\parallel}$ is the projected Jacobian. Substituting this decomposition into the right-hand side of \eqref{eq:standard_lie_alt} and expanding:
\begin{equation}
MH^{(2)} + H^{(2)}M^T = \widetilde{M}H^{(2)} + H^{(2)}\widetilde{M}^T + R, \label{eq:lie_expanded}
\end{equation}
where $R$ collects all terms involving at least one factor of $(I - P_{\parallel})$:
\begin{equation}
\begin{split}
R &= \bigl[P_{\parallel} M (I - P_{\parallel}) + (I - P_{\parallel}) M P_{\parallel} + (I - P_{\parallel}) M (I - P_{\parallel})\bigr] H^{(2)}\\ &+ H^{(2)} \bigl[(I - P_{\parallel}) M^T P_{\parallel} + P_{\parallel} M^T (I - P_{\parallel}) + (I - P_{\parallel}) M^T (I - P_{\parallel})\bigr].
\end{split} \label{eq:R_definition}
\end{equation}

The degeneracy constraint $H^{(2)}\nabla H^{(0)} = 0$ implies that $H^{(2)}$ maps the normal direction $\nabla H^{(0)}$ to zero. Since $(I - P_{\parallel})$ is the projector onto the span of $\nabla H^{(0)}$, we have:
\begin{equation}
H^{(2)}(I - P_{\parallel}) = 0, \qquad (I - P_{\parallel}) H^{(2)} = 0, \label{eq:H2_projector_kill}
\end{equation}
where the second identity follows from the symmetry of $H^{(2)}$: $(I - P_{\parallel}) H^{(2)} = \bigl(H^{(2)}(I - P_{\parallel})\bigr)^T = 0^T = 0$. Applying \eqref{eq:H2_projector_kill} to eliminate terms in $R$, any term containing $(I - P_{\parallel})$ adjacent to $H^{(2)}$ vanishes. Specifically:
\begin{align}
(I - P_{\parallel}) M P_{\parallel} \cdot H^{(2)} &= (I - P_{\parallel}) M \cdot P_{\parallel} H^{(2)} = (I - P_{\parallel}) M \cdot H^{(2)}, \notag \\ (I - P_{\parallel}) M (I - P_{\parallel}) \cdot H^{(2)} &= 0, \notag \\ P_{\parallel} M (I - P_{\parallel}) \cdot H^{(2)} &= 0. \label{eq:R_simplify}
\end{align}
Similarly for the transposed terms. The only surviving contributions in $R$ are:
\begin{equation}
R = (I - P_{\parallel}) M H^{(2)} + H^{(2)} M^T (I - P_{\parallel}) = \Delta[H^{(2)}], \label{eq:R_equals_Delta}
\end{equation}
where we have identified the \emph{rotational compensator}, the linear operator acting on symmetric second-order tensors defined by
\begin{equation}
\Delta[H^{(2)}] := H^{(2)} M^T (I - P_{\parallel}) + (I - P_{\parallel}) M H^{(2)}.
\label{eq:rotational_compensator}
\end{equation}
The structure of $\Delta$ is most transparent in two limiting regimes. When $H^{(0)}=c$ (constant), $P_{\parallel}=I$ and $\Delta$ vanishes identically. When $\nabla H^{(0)} \neq 0$, writing $I - P_{\parallel} = n \otimes n$ with $n = \nabla H^{(0)}/|\nabla H^{(0)}|$, the compensator becomes $\Delta[H^{(2)}] = H^{(2)} M^T n \otimes n + n \otimes n M H^{(2)}$, which is a symmetric rank-at-most-$2$ correction that tracks the rotation of the constraint direction $n$.

Substituting \eqref{eq:R_equals_Delta} back into \eqref{eq:lie_expanded}, the standard Lie transport equation \eqref{eq:standard_lie_alt} becomes:
\begin{equation}
\frac{DH^{(2)}}{Dt} = \widetilde{M}H^{(2)} + H^{(2)}\widetilde{M}^T + \Delta[H^{(2)}], \label{eq:projected_derived}
\end{equation}
which is the projected Lie transport equation. Writing the total derivative explicitly, for $N=2$ truncation it reads
\begin{equation}
\frac{DH^{(2)}}{Dt} := \frac{\partial H^{(2)}}{\partial t} + H^{(1)l}(t,y) \frac{\partial H^{(2)}}{\partial y^l} = \widetilde{M} H^{(2)}  + H^{(2)}\widetilde{M}^T  + \Delta[H^{(2)}]
\label{eq:projected_lie_transport}
\end{equation}
where $\widetilde{M} = P_{\parallel} M P_{\parallel}$ is the projected Jacobian matrix, $P_{\parallel}$ is the orthogonal projector onto the tangent space of the level sets of $H^{(0)}$ (satisfying $P_{\parallel} \nabla H^{(0)} = 0$), and $\Delta[H^{(2)}]$ is the rotational compensator of Eq.~\eqref{eq:rotational_compensator}. The two terms on the right-hand side have distinct geometric origins: $\widetilde{M}H^{(2)} + H^{(2)}\widetilde{M}^T$ is the tangential Lie drag of $H^{(2)}$ along the level sets, while $\Delta[H^{(2)}]$ corrects for the rotation of the constraint direction $n$ as the flow tilts the level sets of $H^{(0)}$. This equation strictly preserves the degeneracy constraint $H^{(2)} \nabla H^{(0)} = 0$ for all $t>0$, provided the initial data satisfy it.

To verify that \eqref{eq:projected_derived} preserves $H^{(2)}\nabla H^{(0)} = 0$, we act on $\nabla H^{(0)}$ from the right. Since $H^{(0)}$ satisfies the zero-order transport equation $\frac{DH^{(0)}}{Dt} = 0$, the gradient evolves as $\frac{D}{Dt}\nabla H^{(0)} = -M^T \nabla H^{(0)}$. Therefore:
\begin{align}
\frac{D}{Dt}\bigl(H^{(2)}\nabla H^{(0)}\bigr) 
&= \frac{DH^{(2)}}{Dt}\nabla H^{(0)} + H^{(2)}\frac{D}{Dt}\nabla H^{(0)} \notag \\
&= \frac{DH^{(2)}}{Dt}\nabla H^{(0)} - H^{(2)}M^T\nabla H^{(0)}. 
\label{eq:constraint_evolution_alt}
\end{align}
Evaluating the first term using \eqref{eq:projected_derived}:
\begin{align}
\frac{DH^{(2)}}{Dt}\nabla H^{(0)} 
&= \widetilde{M}H^{(2)}\nabla H^{(0)} + H^{(2)}\widetilde{M}^T\nabla H^{(0)} + \Delta[H^{(2)}]\nabla H^{(0)} \notag \\ 
&= 0 + H^{(2)}\widetilde{M}^T\nabla H^{(0)} + H^{(2)}M^T\nabla H^{(0)}, \label{eq:first_term_eval}
\end{align}
where we used $H^{(2)}\nabla H^{(0)} = 0$ (killing the first term), and $\Delta[H^{(2)}]\nabla H^{(0)} = H^{(2)}M^T(I - P_{\parallel})\nabla H^{(0)} = H^{(2)}M^T\nabla H^{(0)}$ (using $(I - P_{\parallel})\nabla H^{(0)} = \nabla H^{(0)}$). For the middle term, since $P_{\parallel}\nabla H^{(0)} = 0$:
\begin{equation}
H^{(2)}\widetilde{M}^T\nabla H^{(0)} = H^{(2)}P_{\parallel}M^TP_{\parallel}\nabla H^{(0)} = 0. \label{eq:middle_term_zero}
\end{equation}
Substituting \eqref{eq:first_term_eval} and \eqref{eq:middle_term_zero} into \eqref{eq:constraint_evolution_alt}:
\begin{equation}
\frac{D}{Dt}\bigl(H^{(2)}\nabla H^{(0)}\bigr) = \bigl(0 + 0 + H^{(2)}M^T\nabla H^{(0)}\bigr) - H^{(2)}M^T\nabla H^{(0)} = 0. \label{eq:constraint_preserved_alt}
\end{equation}
Hence, the degeneracy constraint is strictly preserved for all $t > 0$.

When $H^{(0)} = c$ (constant), $\nabla H^{(0)} = 0$ implies $P_{\parallel} = I$, so $\widetilde{M} = M$ and $\Delta[H^{(2)}] = H^{(2)}M^T \cdot 0 + 0 \cdot MH^{(2)} = 0$. Equation \eqref{eq:projected_derived} reduces identically to the standard Lie transport equation. The projected framework is therefore a genuine extension of the standard one: it coincides with it in the constant case and enforces the degeneracy constraint in the non-constant case.

The projector $P_{\parallel}$ enforces a physical selection rule. When $\nabla H^{(0)} \neq 0$, the second-order contact stiffness $H^{(2)}$ is only permitted in directions tangent to the level sets of $H^{(0)}$. The normal direction, corresponding to motion along the gradient of $H^{(0)}$, is ``frozen out'' because the probability field cannot focus across a level set of $H^{(0)}$. The case $\nabla H^{(0)} = 0$ is the free limit where all directions are unfrozen ($P_{\parallel}=I$).

In the case of $\nabla H^{(0)}\neq 0$, the projector $P_{\parallel}$ then removes the single normal direction $\nabla H^{(0)}$, leaving an $(m-1)$-dimensional subspace in which $H^{(2)}$ evolves freely. The resulting $H^{(2)}$ is singular (rank at most $m-1$), which is the geometric signature of a non-constant zero-order contact potential. This rank deficiency is not a defect; it is the contact-geometric encoding of the fact that probability cannot resolve the value of the generalised integral $I(t,y)=H^{(0)}(t,y)$.

\subsection{The Exact Closure of Contact Potential}\label{sec:Truncation}

The master equation~\eqref{eq:master_equation} governs the contact potential $H(t,y,\phi)$, and its exact satisfaction is equivalent to the strict conservation of the constraint function $\varepsilon$ along the contact Hamiltonian flow. A truncated potential $H = \sum_{n=0}^N H^{(n)}$ cannot in general satisfy this equation exactly, because the truncation discards the higher-order terms $H^{(N+1)}, H^{(N+2)}, \dots$ that couple back into the lower orders through the Poisson brackets. 

The question we address here is: under what conditions on the retained coefficients $H^{(0)}, H^{(2)}, \dots, H^{(N)}$ does the truncated potential nevertheless satisfy the master equation exactly, so that no information leaks through the truncation boundary?

We expand the master equation order by order in the fibre coordinate $\phi$. Collecting terms of homogeneous degree $p$, the master equation is equivalent to the infinite sequence of consistency equations $C_p = 0$, where
\[
C_p := (p-1) \frac{\partial H^{(p)}}{\partial t} + \sum_{n=0}^{p+1} (n-1) \{H^{(n)}, H^{(p+1-n)}\} = 0.
\]
With the truncation convention $H^{(N+1)} \equiv 0$ and $H^{(n)} \equiv 0$ for all $n>N$, the time-derivative term vanishes identically for $p \geq N+1$, and the problem reduces to identifying the conditions under which every $C_p$ vanishes.

For orders $p=0$ and $p=1$, we have $C_0 = -\partial_t H^{(0)} - \{H^{(0)}, H^{(1)}\}$. Since $H^{(1)} = H^{(1)i} \phi_i$, the Poisson bracket evaluates to $\{H^{(0)}, H^{(1)}\} = H^{(1)i} \partial_{y^i} H^{(0)}$. Thus $C_0 = 0$ reduces to the zero-order transport condition
$$\frac{\partial H^{(0)}}{\partial t} + H^{(1)i}(t,y) \frac{\partial H^{(0)}}{\partial y^i} = 0,$$
which states that $H^{(0)}$ is Lie-transported along $H^{(1)}$. For $p=1$, we have $C_1 = -2\{H^{(0)}, H^{(2)}\} = -2 H^{(2)ij} \phi_j \partial_{y^i} H^{(0)}$. For this to vanish identically for all $\phi$, the coefficient of each $\phi_j$ must vanish, yielding
$$H^{(2) ij}(t,y) \frac{\partial H^{(0)}}{\partial y^j} = 0, \quad \forall i=1,2,\dots,m.$$
This is the first-order degeneracy condition, the geometric content of which is that $H^{(2)}$ is confined to the tangent space of the level sets of $H^{(0)}$.

For orders $2 \leq p \leq N$, expanding the summation in $C_p$ for $n=0$ and $n=p+1$, and using the antisymmetry of the Poisson bracket $\{H^{(p+1)}, H^{(0)}\} = -\{H^{(0)}, H^{(p+1)}\}$, yields the boundary term $-(p+1)\{H^{(0)}, H^{(p+1)}\}$. The $n=1$ term vanishes because the prefactor is zero. The $n=p$ term gives $(p-1)\{H^{(p)}, H^{(1)}\}$. Thus, $C_p$ simplifies to
\[
C_p = (p-1) \frac{\partial H^{(p)}}{\partial t} + (p-1) \{H^{(p)}, H^{(1)}\} + \sum_{n=2}^{p-1} (n-1) \{H^{(n)}, H^{(p+1-n)}\} - (p+1) \{H^{(0)}, H^{(p+1)}\}.
\]
For this to vanish for every $p=2,3,\dots,N$, the coefficients must satisfy the high-order recurrence condition
\begin{equation*}\begin{split}
(p-1) \frac{\partial H^{(p)}}{\partial t} &+ (p-1) \{H^{(p)}, H^{(1)}\}\\
&+ \sum_{n=2}^{p-1} (n-1) \{H^{(n)}, H^{(p+1-n)}\} \\
&- (p+1) \{H^{(0)}, H^{(p+1)}\} = 0.
\end{split}\end{equation*}
This is a recurrence: each $H^{(p)}$ is determined by the lower-order coefficients $H^{(0)}, \dots, H^{(p-1)}$ together with the next-order coefficient $H^{(p+1)}$, the latter acting as a source term through the bracket $\{H^{(0)}, H^{(p+1)}\}$.

For orders $p \geq N+1$, the time derivative vanishes because $H^{(p)} = 0$. By the truncation convention $H^{(k)} = 0$ for $k > N$, the summation in $C_p$ reduces to indices where both $n \leq N$ and $p+1-n \leq N$. This requires $n \geq p+1-N$. Thus,
\[
C_p = \sum_{n=\max(0, p+1-N)}^{N} (n-1) \{H^{(n)}, H^{(p+1-n)}\}.
\]
These are the residual Poisson brackets that survive the truncation: they couple the retained coefficients $H^{(n)}$ with $n \leq N$ to the discarded coefficients $H^{(p+1-n)}$ with $p+1-n \geq p+1-N \geq 1$. For $C_p$ to vanish for all $p \geq N+1$, these residual brackets must vanish identically, yielding the structural closure condition
\begin{equation*}
\sum_{n=\max(0, p+1-N)}^{N} (n-1) \{H^{(n)}, H^{(p+1-n)}\} = 0.
\end{equation*}

Collecting the four requirements that emerge from the order-by-order analysis, the truncated contact potential $H = \sum_{n=0}^N H^{(n)}$ exactly satisfies the master equation \eqref{eq:master_equation} if and only if $H^{(0)}, H^{(2)}, \dots, H^{(N)}$ satisfy: 
\begin{enumerate}[label=(\roman*)]
\item the zero-order transport condition; 
\item the first-order degeneracy condition;
\item the high-order recurrence condition for all $p=2,3,\dots,N$;
\item the structural closure condition for all $p \geq N+1$, with the truncation convention $H^{(N+1)} \equiv 0$ and $H^{(n)} \equiv 0$ for all $n>N$. 
\end{enumerate}

When these conditions hold, the constraint function $\varepsilon$ is strictly conserved along the contact Hamiltonian flow.

For the physically fundamental case of $N=2$ truncation, which is the focus of this paper, the structural closure condition (iv) is automatically satisfied. For $N=2$ and $p \geq 3$, the lower bound of the sum is $p+1-2 = p-1 \geq 2$. Since the upper bound is $N=2$, the only possible term occurs when $p=3, n=2$, giving the single bracket $\{H^{(2)}, H^{(2)}\}$, which vanishes identically by the antisymmetry of the Poisson bracket. For $p > 3$, the summation interval is empty. Thus for $N=2$, conditions (i)--(iii) are sufficient to guarantee the exact satisfaction of the master equation, and no additional closure assumption is needed. Moreover, when $H^{(0)}=c$ is constant, condition (ii) is automatically satisfied and the high-order recurrence condition simplifies; in particular, for $N=2$ truncation, it reduces to the Lie transport equation \eqref{eq:second_order_lie_transport}.

\subsection{Zero-Order Contact Potential $H^{(0)}$}\label{sec:H0_construction}
 
The zero-order contact potential is determined by solving the zero-order transport equation \eqref{eq:zero_order_transport}, a first-order linear partial differential equation. We derive its general solution---the master formula---by a separated ansatz that reduces the transport equation to a defining relation between a geometric potential $S$ and a source $\alpha$. The specific form of $H^{(0)}$ for each system type is then obtained by fixing $S$ and $\alpha$ through the invariant-measure construction, which selects the ergodic invariant measure supported on the $\omega$-limit set of the flow generated by $H^{(1)}$.

\subsubsection{The Transport Equation of $H^{(0)}$ and General Solution}
Let $\mathcal{L} := H^{(1)i} \partial/\partial y^i$ denote the Lie derivative along the vector field $H^{(1)i}(t,y)$, and let $\Phi_t$ denote the flow generated by $H^{(1)}$. The flow $\{\Phi_t\}_{t \in \mathbb{R}}$ constitutes a one-parameter group of diffeomorphisms acting on the base manifold $E$. The zero-order transport equation \eqref{eq:zero_order_transport} demands that $H^{(0)}$ is invariant under the extended flow $\partial_t + \mathcal{L}$:
\begin{equation}
\frac{\partial H^{(0)}}{\partial t} + \mathcal{L} H^{(0)} = 0.
\label{eq:H0_transport_recast}
\end{equation}
This equation states that $H^{(0)}$ is a first integral of the extended flow. To solve it systematically, we introduce a smooth geometric potential $S(y)$ and define its associated source $\alpha(y) := \mathcal{L} S(y)$. The pair $(S, \alpha)$ will be fixed subsequently by the invariant-measure construction. The defining relation
\begin{equation}
\mathcal{L} S(y) = \alpha(y).
\label{eq:source_relation}
\end{equation}
reduces the problem of determining $H^{(0)}$ to the identification of the geometric potential $S$ and its source $\alpha$.

The defining relation \eqref{eq:source_relation} allows us to seek a separated solution of the transport equation \eqref{eq:H0_transport_recast} of the form $H^{(0)}(t, y) = S(y) + F(t, y)$. Substituting this ansatz and using \eqref{eq:source_relation} yields $\partial_t F + \mathcal{L} F = -\alpha(y)$. To satisfy this, we construct $F$ by accumulating the source $\alpha$ along the backward characteristic $\Phi_{-\sigma}(y)$. Setting the initial condition $F(0, y) = 0$, the explicit time-dependence is given by the integral:
\begin{equation}
F(t, y) = -\int_0^t \alpha\bigl(\Phi_{-\sigma}(y)\bigr)\,d\sigma.
\label{eq:F_integral}
\end{equation}
To verify this construction, we compute the material derivative of $F$. Using the chain rule and the identity $\frac{d}{d\sigma}\alpha(\Phi_{-\sigma}(y)) = -(\mathcal{L} \alpha)(\Phi_{-\sigma}(y))$, we find $\mathcal{L} F = \alpha(\Phi_{-t}(y)) - \alpha(y)$. The partial time derivative gives $\partial_t F = -\alpha(\Phi_{-t}(y))$. Summing these yields $\partial_t F + \mathcal{L} F = -\alpha(y)$, which exactly cancels the $\alpha(y)$ from $\mathcal{L} S$. Thus, 
\begin{equation}
H^{(0)}(t, y) = S(y) - \int_0^t \alpha\bigl(\Phi_{-\sigma}(y)\bigr)\,d\sigma
\label{eq:master_formula}
\end{equation}
strictly satisfies the zero-order transport equation. Crucially, the verification never uses the assumption that $\alpha$ is constant; it is valid for arbitrary source functions $\alpha(y)$, whether state-dependent or not. This generality is the mathematical root of the method's universality: it applies to all system types without breakdown, and the system's topology enters only through the specific choice of $S$ and $\alpha$.

\subsubsection{The Invariant-Measure Construction}

The master formula~\eqref{eq:master_formula} expresses $H^{(0)}$ through a geometric potential $S$ and a source $\alpha$ linked by the defining relation $\mathcal{L} S = \alpha$. We fix both objects by the following construction.

Let $\Omega$ denote the canonical volume form induced on the base manifold by the coordinates $y^i$ (equivalently, the contact volume form $\Theta\wedge(d\Theta)^n$ restricted to the base). Let $\mu$ denote the ergodic invariant measure of the flow $\Phi_t$ generated by $H^{(1)}$ supported on the $\omega$-limit set of a typical trajectory. Whenever $\mu$ is absolutely continuous with respect to $\Omega$ on the support of interest, denote its Radon--Nikodym derivative by
\begin{equation}
\rho(y) := \frac{d\mu}{d\Omega}(y).
\end{equation}
The source and the geometric potential are then defined by
\begin{equation}
\alpha(y) := -\,\mathcal{L} \ln \Omega, \qquad S(y) := \ln \rho(y) = \ln \frac{d\mu}{d\Omega}(y).
\label{eq:invariant_measure_rule}
\end{equation}

The invariance $\Phi_t^*\mu = \mu$ reads, in density form, $\mathcal{L}(\rho\,\Omega)=0$. Expanding by the Leibniz rule,
\begin{equation}
\mathcal{L}(\rho\,\Omega) = (\mathcal{L} \rho)\,\Omega + \rho\,(\mathcal{L}\Omega) = 0
\;\Longrightarrow\;
\mathcal{L} \ln \rho = -\,\mathcal{L} \ln \Omega.
\label{eq:continuity_log_form}
\end{equation}
The left-hand side is $\mathcal{L} S$ and the right-hand side is $\alpha$, so the defining relation
\begin{equation}
\mathcal{L} S = \alpha
\label{eq:source_from_invariance}
\end{equation}
is the continuity equation of the invariant measure in logarithmic form. Substituting~\eqref{eq:invariant_measure_rule} into the master formula gives
\begin{equation}
H^{(0)}(t,y) = \ln\frac{d\mu}{d\Omega}(y) + \int_0^t \mathcal{L}\ln\Omega\bigl(\Phi_{-\sigma}(y)\bigr)\,d\sigma.
\label{eq:H0_universal}
\end{equation}

\subsubsection{Four Typical Regimes }

The four typical regimes correspond to four forms of the ergodic invariant measure $\mu$.

\textbf{Point attractor.} For a trajectory $y(t)\to y_*$ converging to a hyperbolic fixed point $y_*$, the $\omega$-limit set is $\{y_*\}$ and the ergodic invariant measure is the Dirac measure $\mu = \delta_{y_*}$, invariant since $\Phi_t(y)\to y_*$ implies $\Phi_t^*\delta_{y_*}=\delta_{y_*}$. The density $\rho = d\delta_{y_*}/d\Omega$ is singular, so $S=\ln\rho$ is defined through a regularising family $\mu_\varepsilon$ of smooth invariant measures (for instance, the stationary measures of a vanishing-noise Fokker--Planck semigroup centred at $y_*$), with $\mu_\varepsilon\to\delta_{y_*}$ weakly as $\varepsilon\to 0$. Along any backward characteristic $\Phi_{-\sigma}(y)$ that diverges from $y_*$, the integral $\int_0^t \mathcal{L}\ln\Omega\,d\sigma$ diverges to $+\infty$ at the same rate at which $\ln\rho_\varepsilon(y)\to-\infty$ off $y_*$; the two divergences cancel in~\eqref{eq:H0_universal}, and the regularised potentials $H^{(0)}_\varepsilon$ converge to a finite constant. Absorbing this constant into the normalisation,
\begin{equation}
H^{(0)} = c.
\label{eq:H0_point}
\end{equation}

\textbf{Limit cycle.} For a periodic orbit $\gamma$ of period $T$, the $\omega$-limit set is $\gamma$ and the ergodic invariant measure $\mu_\gamma$ is the equitime measure on $\gamma$, that is, the probability measure uniformly distributed in time along the cycle. It is invariant by construction: $\Phi_t^*\mu_\gamma=\mu_\gamma$ for all $t$. Its density $\rho_\gamma = d\mu_\gamma/d\Omega$ is supported on $\gamma$ and is a smooth periodic function of arc length along the cycle. The construction~\eqref{eq:invariant_measure_rule} gives
\begin{equation}
S(y) = \ln\rho_\gamma(y), \qquad \alpha(y) = -\,\mathcal{L}\ln\Omega = -\,\Lambda(y),
\end{equation}
and the master formula yields
\begin{equation}
H^{(0)}(t,y) = \ln\rho_\gamma(y) + \int_0^t \Lambda\bigl(\Phi_{-\sigma}(y)\bigr)\,d\sigma.
\label{eq:H0_limit_cycle_unified}
\end{equation}
On the cycle, $\Lambda$ integrates over one period to the transverse contraction rate, and the level sets of $H^{(0)}$ coincide with the isochrons of the cycle. In the basin of attraction, $\rho_\gamma$ is extended off the cycle by forward-orbit convergence: every point $y$ in the basin satisfies $\Phi_t(y)\to\gamma$, so the forward limit of $\rho_\gamma$ along the orbit defines the basin value. The angular structure of the cycle is encoded in the spatial dependence of $\rho_\gamma$.

\textbf{Chaos.} For a typical trajectory on a strange attractor $A$, the $\omega$-limit set is $A$ itself and the ergodic invariant measure is the Sinai--Ruelle--Bowen (SRB) measure $\mu_{\mathrm{SRB}}$, selected by Lebesgue-almost-every initial condition in the basin. Its density $\rho_{\mathrm{SRB}} = d\mu_{\mathrm{SRB}}/d\Omega$ exists along the unstable directions of the flow and satisfies $\mathcal{L}(\rho_{\mathrm{SRB}}\Omega)=0$ by invariance. The construction~\eqref{eq:invariant_measure_rule} gives
\begin{equation}
S(y) = \ln\rho_{\mathrm{SRB}}(y), \qquad \alpha(y) = -\,\Lambda(y),
\end{equation}
and the master formula becomes
\begin{equation}
H^{(0)}(t,y) = \ln\rho_{\mathrm{SRB}}(y) + \int_0^t \Lambda\bigl(\Phi_{-\sigma}(y)\bigr)\,d\sigma.
\label{eq:H0_chaos_unified}
\end{equation}
When the divergence is state-independent, $\Lambda(y)\equiv\Lambda_0<0$, the integral evaluates to $\Lambda_0 t$ and~\eqref{eq:H0_chaos_unified} reduces to $H^{(0)}(t,y)=\ln\rho_{\mathrm{SRB}}(y)+\Lambda_0 t$. When the divergence is state-dependent, the full integral is retained and $\alpha(y)=-\Lambda(y)$ is a local quantity.

\textbf{Conservative system.} For a quasi-periodic trajectory on a KAM torus (or, in the integrable limit, on an invariant torus of the action--angle decomposition), the ergodic invariant measure is the Liouville measure $\mu_{\mathrm{L}}$ restricted to the torus, invariant by Hamilton's theorem: $\mathcal{L}\Omega=0$, so $\Phi_t^*\mu_{\mathrm{L}}=\mu_{\mathrm{L}}$. The construction~\eqref{eq:invariant_measure_rule} gives
\begin{equation}
\alpha(y) = -\,\mathcal{L}\ln\Omega = 0, \qquad S(y) = \ln\frac{d\mu_{\mathrm{L}}}{d\Omega}(y).
\end{equation}
Since the Liouville measure is the measure induced by $\Omega$ on phase space, $d\mu_{\mathrm{L}}/d\Omega \equiv 1$, and therefore $S\equiv 0$. The master formula reduces to
\begin{equation}
H^{(0)} = c,
\label{eq:H0_conservative}
\end{equation}
with the constant fixed by normalisation. Deterministic focusing does not occur because the projected Jacobian lacks negative real eigenvalues.

Table~\ref{tab:unified_H0} summarises the four regimes.
\begin{table}[H]
\centering
\caption{The four typical regimes as specialisations of the invariant-measure construction~\eqref{eq:invariant_measure_rule}.}
\label{tab:unified_H0}
\small
\renewcommand{\arraystretch}{1.45}
\newcolumntype{L}[1]{>{\RaggedRight\arraybackslash}p{#1}}
\newcolumntype{M}[1]{>{$\RaggedRight\arraybackslash}p{#1}<{$}}
\begin{tabular}{@{}L{2.2cm}L{2.8cm}L{3.5cm}M{3.0cm}M{3.0cm}@{}}
\toprule
Regime & Representative motion & Ergodic invariant measure $\mu$ & S=\ln(d\mu/d\Omega) & \alpha=-\mathcal{L}\ln\Omega \\
\midrule
Point attractor & $y(t)\to y_*$ & Dirac $\delta_{y_*}$ & \ln\rho\ \text{(regularised)} & -\Lambda(y) \\
Limit cycle & periodic orbit $\gamma$ & equitime measure on $\gamma$ & \ln\rho_\gamma & -\Lambda(y) \\
Chaos & SRB-typical trajectory & SRB measure $\mu_{\mathrm{SRB}}$ & \ln\rho_{\mathrm{SRB}} & -\Lambda(y) \\
Conservative & quasi-periodic (KAM) & Liouville $\mu_{\mathrm{L}}$ & 0 & 0 \\
\bottomrule
\end{tabular}
\end{table}

\subsection{First-Order Contact Potential and the Zero-Fibre Dynamics}
\label{subsec:drift_field}

The contact dynamical equation~\eqref{eq:contact_hamiltonian_standard} establishes that the evolution velocity of the base coordinates is given by $\dot{y}^i = \partial H/\partial \phi_i$. Using the Taylor expansion~\eqref{eq:taylor_expansion_H} of the contact potential, we note that the zero-order term $H^{(0)}$ is independent of $\phi$, and thus its derivative vanishes everywhere. Furthermore, each higher-order term $H^{(n)}$ (for $n \ge 2$) is a homogeneous polynomial of degree $n \ge 2$ in $\phi$; its partial derivative with respect to $\phi_i$ is homogeneous of degree $n-1 \ge 1$, and therefore strictly vanishes at $\phi=0$.

Consequently, along the zero-fibre section, the sole surviving contribution comes from the first-order term:
\begin{equation}
\left.\frac{\partial H}{\partial \phi_i}\right|_{\phi=0} = H^{(1)i}(t,y).
\label{eq:drift_field}
\end{equation}

\begin{equation}
\dot{y}^i\bigg|_{\phi=0} = H^{(1)i}(t,y).
\label{eq:drift_first_order}
\end{equation}

The coefficient $H^{(1)i}$ governs the evolution on the zero-fibre section. Its full physical significance will be established later in Section~\ref{sec:Emergence-Determinism}.

\subsection{Second-Order Contact Potential $H^{(2)}$}

With the zero-order contact potential determined, we now construct the second-order term. For $N=2$ truncation, the second-order tensor $H^{(2)}$ is governed by the unified projected Lie transport equation \eqref{eq:projected_lie_transport}. On the subspace where $H^{(0)}=c$ (dissipative systems or the degenerate directions of any system), this reduces to the standard Lie transport equation.

Along the characteristic line $y(t) = Y(t; 0, y_0)$, the solution to the projected tensor Lie transport equation \eqref{eq:projected_lie_transport} is
\begin{equation}
\begin{split}
H^{(2)}(t, Y(t; 0, y_0)) &= \widetilde{\Phi}(t, 0; y_0) H_0^{(2)}(y_0) \widetilde{\Phi}^T(t, 0; y_0) \\&+ \int_0^t \widetilde{\Phi}(t, \tau; y_0) \Delta[H^{(2)}(\tau)] \widetilde{\Phi}^T(t, \tau; y_0) \, d\tau.
\end{split}
\label{eq:H2_characteristic_solution_general}
\end{equation}
where $H_0^{(2)}(y_0) = H^{(2)}(0, y_0)$ is the initial second-order tensor satisfying $H^{(2)}_0 \nabla H^{(0)}(0,y_0)=0$, $\widetilde{\Phi}(t, s; y_0)$ is the fundamental solution matrix of the projected variational equation $$\dot{\widetilde{\Phi}}(t, s; y_0) = \widetilde{M}(t, Y(t; s, y_0)) \widetilde{\Phi}(t, s; y_0)$$ with $\widetilde{\Phi}(s, s; y_0) = I$, and $\Delta[H^{(2)}(\tau)]$ denotes $\Delta[H^{(2)}](\tau, Y(\tau; 0, y_0))$. This is verified by differentiating \eqref{eq:H2_characteristic_solution_general} along the characteristic line, applying the Leibniz integral rule, substituting the variational equation $\dot{\widetilde{\Phi}} = \widetilde{M}\widetilde{\Phi}$, and factoring to recover $\widetilde{M}H^{(2)} + H^{(2)}\widetilde{M}^T + \Delta[H^{(2)}]$. The initial condition is satisfied since $\widetilde{\Phi}(0,0)=I$ and the integral vanishes at $t=0$.

When $H^{(0)}=c$ (as established for dissipative systems), $P_\parallel = I$, $\widetilde{M} = M$, and $\Delta[H^{(2)}] = 0$. The integral term disappears, and the exact solution reduces to the standard homogeneous Lyapunov form:
\begin{equation*}
H^{(2)}(t, Y(t; 0, y_0)) = \Phi(t, 0; y_0) H_0^{(2)}(y_0) \Phi^T(t, 0; y_0).
\end{equation*}

The second-order contact term $H^{(2)}$ is the key geometric element through which the hidden coupling between probability and determinism becomes visible. Without it, contact dynamics degenerates into a trivial deterministic ordinary differential equation, losing all structural content. Its core roles are as follows. 

First, $H^{(2)}$ realizes bidirectional coupling between the base and fiber coordinates: for $N=1$ truncation without the second-order term, the base evolution is completely independent of $\phi$; with the introduction of $H^{(2)}$, the base evolution becomes $\dot{y}^i = H^{(1)i}(t,y) + H^{(2) ij}(t,y)\phi_j$, where changes in $\phi$ can back-react to the base evolution. 

Second, $H^{(2)}$ is the core carrier of constraint function conservation: for $N=2$ truncation, the constraint function is $$\varepsilon = -H^{(0)} + \frac{1}{2}\phi^T H^{(2)}(t,y) \phi,$$ and $H^{(2)}$ undergoes Lie transport along the characteristic lines of the flow generated by $H^{(1)}$, decaying synchronously to exactly counterbalance the growth of $\phi$. Third, $H^{(2)}$ defines the second-order contact stiffness of fluctuations: its eigenvalues correspond to the second-order contact stiffness in each direction of $\phi$. Fourth, the diagonalisation of $H^{(2)}$ via an orthogonal contact transformation is the core step in decoupling the nonlinear Hamilton-Jacobi equation into multiple one-dimensional equations.

\section{Emergence of Determinism}\label{sec:Emergence-Determinism}

\subsection{The Deterministic Manifold}

The zero-fibre section $\phi=0$ defines a submanifold $\mathcal{S} \subset T^*E \times \mathbb{R}$ of the contact manifold, which serves as the deterministic manifold of the system. On $\mathcal{S}$, the contact Hamiltonian equations reduce to (cf.\ Section~\ref{subsec:drift_field})
\begin{equation}
\dot{y}^i\big|_{\phi=0} = H^{(1)i}(t,y), \quad \dot{\phi}_i\big|_{\phi=0} = -\frac{\partial H^{(0)}}{\partial y^i}. \label{eq:skeleton_dynamics}
\end{equation}
For dissipative systems where $H^{(0)}=c$ (constant), we have $\dot{\phi}_i\big|_{\phi=0} = 0$. Consequently, $\mathcal{S}$ is an invariant submanifold: trajectories starting on the manifold remain on it indefinitely.

The fibre coordinate $\phi_i$ quantifies the spatial structure of the probability field. According to the jet-bundle structure of the contact manifold, $\phi_i$ is a weighted sum of all orders of probability gradients:
\begin{equation*}
\phi_i = \sum_{1\le k \le \infty}\sum_{\mu_1<\cdots<\mu_k}B_i^{\mu_1\cdots\mu_k}P_{\mu_1\cdots\mu_k},
\end{equation*}
where $B_i^{\mu_1\cdots\mu_k}$ is the function of cumulants of stochastic systems. With the coefficients $B_i^{\mu_1\cdots\mu_k}$ being bounded functionals of the system cumulants, the fibre coordinate $\phi_i$ is well defined, and $\phi=0$ signifies that the weighted sum of all spatial probability gradients vanishes identically. This encompasses two fundamental physical scenarios: (i) Equiprobable distribution: $P=\mathrm{const}$, where all spatial derivatives $P_{\mu_1\cdots\mu_k}=0$ for $k \ge 1$, causing the sum to vanish trivially. (ii) Steady-state distribution: A non-uniform distribution where higher-order spatial gradients exist ($P_{\mu_1\cdots\mu_k} \neq 0$ for some $k \ge 1$), but their weighted sum exactly cancels out to zero in the fibre coordinate $\phi_i$. This corresponds to steady-state configurations in which the probability flux is in geometric balance.

Conversely, $\phi \neq 0$ indicates that the system has deviated from these balanced configurations. The growth of $|\phi|$ corresponds to an increasing deviation of the probability field from its steady-state or equiprobable configuration, reflecting a strengthening of the net directional probability gradient. It is worth noting that the zero-fibre section $\phi=0$ contains both the equiprobable state ($P(y)=\mathrm{const}$, where all derivatives vanish trivially) and the steady state (where the weighted sum of derivatives vanishes by geometric balance), but not the delta function $P(y)=\delta(y-y_0)$, whose derivatives do not vanish and which therefore has $\phi \neq 0$. The delta function is the target of gradient amplification, not the starting point.

Although $\mathcal{S}$ is invariant for dissipative systems, the dynamics in its neighbourhood reveals a crucial instability: infinitesimal perturbations with $\phi \neq 0$ undergo exponential amplification. We now examine how the contact constraint resolves this apparent paradox.

\subsection{Gradient Amplification and the Contact Constraint}\label{subsec:contact_constraint}
The linearised evolution of the fibre coordinate near the deterministic manifold $\mathcal S$ is governed by the drift-field Jacobian $M$; for the dissipative case $H^{(0)}=c$, this reduces to
\begin{equation}\label{eq:phi_linearized}
\dot\phi = -M^{T}(t,y(t))\,\phi + O(|\phi|^{2}).
\end{equation}
Because the spectrum of $M$ has strictly negative real parts, the eigenvalues of $-M^{T}$ have strictly positive real parts, and any infinitesimal perturbation $\phi(0)\neq 0$ triggers exponential growth
\begin{equation}\label{eq:gradient_amplification}
|\phi(t)| \sim |\phi(0)|\,e^{\sigma t},\qquad 
\sigma := \max_{k}\bigl(-\mathrm{Re}(\lambda_{k})\bigr)>0,
\end{equation}
with characteristic timescale $\tau_{f}=1/\sigma$.  
If $H^{(2)}$ remained of order one, the back-reaction $H^{(2)}\phi$ in the macroscopic equation would diverge.  
The contact constraint resolves this crisis through two exact identities that rigidly couple the growth of $\phi$ to the decay of $H^{(2)}$.
\subsubsection{The projection identity}
The first-order degeneracy condition \eqref{eq:first_order_degeneracy} states that $\nabla H^{(0)}$ lies in the kernel of $H^{(2)}$.  
Writing it in matrix form and inserting the orthogonal projector $P_{\parallel}$ defined in \eqref{eq:projector} together with its complement $I-P_{\parallel}=\mathbf{n}\otimes\mathbf{n}$, we obtain
\begin{equation}\label{eq:H2_right_kernel}
H^{(2)}(I-P_{\parallel})
= (H^{(2)}\mathbf{n})\mathbf{n}^{T}
= \frac{1}{|\nabla H^{(0)}|}\bigl(H^{(2)}\nabla H^{(0)}\bigr)\mathbf{n}^{T}
= \mathbf{0}.
\end{equation}
By the symmetry of $H^{(2)}$, transposing \eqref{eq:H2_right_kernel} yields
\begin{equation}\label{eq:H2_left_kernel}
(I-P_{\parallel})H^{(2)} = \mathbf{0}.
\end{equation}
Inserting the resolution of the identity $I=P_{\parallel}+(I-P_{\parallel})$ on both sides of $H^{(2)}$ gives
\begin{equation}\label{eq:sandwich_expansion}
\begin{split}
H^{(2)}
&= \bigl[P_{\parallel}+(I-P_{\parallel})\bigr] H^{(2)} \bigl[P_{\parallel}+(I-P_{\parallel})\bigr]\\
&= P_{\parallel}H^{(2)}P_{\parallel} + P_{\parallel}H^{(2)}(I-P_{\parallel}) + (I-P_{\parallel})H^{(2)}P_{\parallel} + (I-P_{\parallel})H^{(2)}(I-P_{\parallel}).
\end{split}
\end{equation}
The last three terms vanish by \eqref{eq:H2_right_kernel} and \eqref{eq:H2_left_kernel}, leaving the projection identity
\begin{equation}\label{eq:H2_projection_identity}
H^{(2)} = P_{\parallel} H^{(2)} P_{\parallel}.
\end{equation}
Thus the image and kernel of $H^{(2)}$ are both confined to the tangent space of the level sets of $H^{(0)}$.  
Defining the tangential fibre coordinate $\phi_{\parallel}:=P_{\parallel}\phi$, the constraint function reduces to
\begin{equation}\label{eq:reduced_constraint}
\varepsilon + H^{(0)} = \tfrac{1}{2}\,\phi_{\parallel}^{\!T} H^{(2)}\phi_{\parallel} = \mathrm{const},
\end{equation}
which depends only on the tangential component of the probability gradient.
\subsubsection{Homogeneous tangential Lie transport}
The projected Lie transport equation \eqref{eq:projected_lie_transport} governs the full second-order tensor, with rotational compensator $\Delta[H^{(2)}]$ defined in \eqref{eq:rotational_compensator}.  
To extract the tangential block, we act with $P_{\parallel}$ from both sides.  
Using \eqref{eq:H2_right_kernel}--\eqref{eq:H2_left_kernel} and the orthogonality $(I-P_{\parallel})P_{\parallel}=\mathbf{0}$, the compensator vanishes identically in the tangential--tangential subspace:
\begin{equation}\label{eq:Delta_tangential_zero}
P_{\parallel}\Delta[H^{(2)}]P_{\parallel}
= P_{\parallel}H^{(2)}M^{T}(I-P_{\parallel})P_{\parallel} + P_{\parallel}(I-P_{\parallel})MH^{(2)}P_{\parallel}
= \mathbf{0}.
\end{equation}
For the remaining terms, the idempotence $P_{\parallel}\widetilde{M}=\widetilde{M}$ and $\widetilde{M}^{T}P_{\parallel}=\widetilde{M}^{T}$ (with $\widetilde{M}=P_{\parallel}MP_{\parallel}$ from \eqref{eq:projected_jacobian}), together with \eqref{eq:H2_projection_identity}, yield
\begin{equation}\label{eq:tangential_terms_simplification}
P_{\parallel}\bigl(\widetilde{M}H^{(2)}\bigr)P_{\parallel} = \widetilde{M}H^{(2)},
\qquad
P_{\parallel}\bigl(H^{(2)}\widetilde{M}^{T}\bigr)P_{\parallel} = H^{(2)}\widetilde{M}^{T}.
\end{equation}
Hence, for any tangent vectors $v,w$, the tangential block satisfies the homogeneous equation
\begin{equation}\label{eq:tangential_block_homogeneous}
v^{T}\frac{DH^{(2)}}{Dt}w
= v^{T}\Bigl(\widetilde{M}H^{(2)} + H^{(2)}\widetilde{M}^{T}\Bigr)w.
\end{equation}
Since the projection identity \eqref{eq:H2_projection_identity} implies that $H^{(2)}$ lies entirely within the tangential--tangential block, we denote the tangential block as $H^{(2)}_{\parallel}=P_{\parallel}H^{(2)}P_{\parallel}=H^{(2)}$. Its component-wise evolution within the tangent space corresponds to the standard Lyapunov equation:
\begin{equation}\label{eq:lyapunov_tangential}
\frac{DH^{(2)}_{\parallel}}{Dt}
= \widetilde{M}H^{(2)}_{\parallel} + H^{(2)}_{\parallel}\widetilde{M}^{T},
\end{equation}
where the derivative on the left-hand side is understood as the evolution of the tangential--tangential block of the tensor, as extracted in \eqref{eq:tangential_block_homogeneous}.
\subsubsection{Decay of the effective coupling}
The two identities jointly determine the asymptotic fate of the macroscopic--microscopic coupling.  
From \eqref{eq:linearized_fibre}, the tangential component $\phi_{\parallel}$ grows exponentially at rate $\sigma_{\parallel}:=\min_{k}\bigl(-\mathrm{Re}(\lambda_{k}^{\widetilde{M}})\bigr)>0$:
\begin{equation}\label{eq:phi_perp_growth}
|\phi_{\parallel}(t)| \sim e^{\sigma_{\parallel}t}.
\end{equation}
Meanwhile, because the eigenvalues of $\widetilde{M}$ have strictly negative real parts, the Lyapunov equation \eqref{eq:lyapunov_tangential} forces the tangential stiffness to decay as
\begin{equation}\label{eq:H2_decay}
\|H^{(2)}(t)\| \sim e^{-2\sigma_{\parallel}t}.
\end{equation}
The reduced constraint \eqref{eq:reduced_constraint} rigidly locks these two processes in exact opposition: the product $\phi_{\parallel}^{T}H^{(2)}\phi_{\parallel}$ must remain constant.  
Consequently, the effective coupling vanishes exponentially:
\begin{equation}\label{eq:effective_coupling_decay}
\bigl|H^{(2)}(t)\,\phi_{\parallel}(t)\bigr|
\leq \|H^{(2)}(t)\|\,|\phi_{\parallel}(t)|
\sim e^{-2\sigma_{\parallel}t}\,e^{\sigma_{\parallel}t}
= e^{-\sigma_{\parallel}t}
\;\longrightarrow\; 0.
\end{equation}
The system becomes internally finer ($|\phi_{\parallel}|\to\infty$, the probability field sharpens) while externally blinder ($H^{(2)}\to 0$, the coupling softens), so that the macroscopic variable $y$ evolves as if the internal probability structure did not exist.  
The macroscopic equation therefore converges strictly onto the deterministic drift $\dot y^{i}=H^{(1)i}(t,y)$.  
The contact constraint has forced the effective coupling to vanish, and determinism emerges as a geometric attractor of the contact flow.  
The full variational structure of this deterministic focusing, including the Jacobi-field interpretation and the two distinct focusing regimes, is developed in the next subsection.

\subsection{Deterministic Focusing as a Geometric Attractor}\label{subsec:deterministic_focusing}

\subsubsection{Definition and Variational Equation of the Jacobi Field}

The mechanism by which the contact constraint forces the effective coupling to vanish takes on a transparent geometric meaning when expressed in terms of Jacobi fields of the flow generated by $H^{(1)}$. Defining
\begin{equation}\label{eq:jacobi_field_def}
\xi^i(t,y,\phi) := H^{(2)ij}(t,y)\,\phi_j = y^i - H^{(1)i},
\end{equation}
one verifies that $\xi$ is an exact Jacobi field of the flow generated by $H^{(1)}$. To see this, compute its material derivative along the flow:
\begin{equation*}
\frac{D\xi^i}{Dt} = \frac{DH^{(2)ij}}{Dt}\phi_j + H^{(2)ij}\frac{D\phi_j}{Dt}.
\end{equation*}

For $H^{(0)}=c$, the standard Lie transport \eqref{eq:second_order_lie_transport} gives $\frac{DH^{(2)}}{Dt}=MH^{(2)}+H^{(2)}M^T$, and the linearised fibre equation \eqref{eq:phi_linearized} with $\nabla H^{(0)}=0$ gives $\frac{D\phi}{Dt}=-M^T\phi$. Substituting these, the terms $H^{(2)}M^T\phi$ cancel exactly, yielding
\begin{equation}\label{eq:jacobi_equation}
\frac{D\xi}{Dt} = M \xi .
\end{equation}

For $H^{(0)}\neq c$, the projected Lie transport \eqref{eq:projected_lie_transport} replaces the standard transport. Substituting 
\begin{equation*}
 \frac{DH^{(2)}}{Dt}= \widetilde{M}H^{(2)} + H^{(2)}\widetilde{M}^T + \Delta[H^{(2)}]
\end{equation*}
and 
\begin{equation*}
\frac{D\phi}{Dt} = -\nabla H^{(0)} - M^T\phi
\end{equation*} 
into
\begin{equation*}
\frac{D\xi}{Dt} = (DH^{(2)}/Dt)\phi + H^{(2)}(D\phi/Dt),
\end{equation*}
 the term $-H^{(2)}\nabla H^{(0)}$ vanishes by the first-order degeneracy condition \eqref{eq:first_order_degeneracy}. 
 
The difference 
\begin{equation*}
H^{(2)}(\widetilde{M}^T - M^T)\phi = -H^{(2)}M^T(I-P_\parallel)\phi,
\end{equation*}
using $H^{(2)}=H^{(2)}P_\parallel$ and $\widetilde{M}^T=P_\parallel M^T P_\parallel$, is exactly cancelled by the corresponding part of the compensator 
\begin{equation*}
\Delta[H^{(2)}]\phi = H^{(2)}M^T(I-P_\parallel)\phi + (I-P_\parallel)M\xi
\end{equation*}
leaving 
\begin{equation}
\frac{D\xi}{Dt }= \widetilde{M}\xi + (I-P_\parallel)M\xi.
\label{eq:jacobi_equation_c}
\end{equation}
For $H^{(0)}=c$ this reduces to \eqref{eq:jacobi_equation}; in either case, $\xi$ satisfies a variational equation whose tangential projection is developed below.

By the same degeneracy condition, $n\cdot \xi=0$ for $n=\nabla H^{(0)}/|\nabla H^{(0)}|$, so $\xi$ is strictly tangential to the level sets of $H^{(0)}$:
\begin{equation}\label{eq:u_tangential_new}
\xi(t) \in (\nabla H^{(0)})^{\parallel} \quad\text{for all }t .
\end{equation}
To make the tangential propagation explicit, project the Jacobi equation \eqref{eq:jacobi_equation} onto the tangent and normal subspaces. Decompose the Jacobian as
\begin{equation*}
M = P_\parallel M P_\parallel + P_\parallel M (I-P_\parallel) + (I-P_\parallel) M P_\parallel + (I-P_\parallel) M (I-P_\parallel).
\end{equation*}
Acting on $\xi$ and using $P_\parallel \xi = \xi$ together with $(I-P_\parallel)\xi=0$, the terms involving $(I-P_\parallel)$ reduce to a single normal reaction, giving
\begin{equation}\label{eq:projected_jacobi}
\frac{D\xi}{Dt} = \tilde{M}\xi - n(\dot{n}\cdot \xi) ,
\end{equation}
where $\tilde{M}\xi=(P_\parallel M P_\parallel)\xi$ is the intrinsic tangential propagation along the level set, and $-n(\dot{n}\cdot \xi)$ is the normal constraint reaction that preserves $n\cdot \xi=0$ as the level set rotates. Crucially, the normal fibre component $\phi_n n$ does not appear in \eqref{eq:projected_jacobi}: since $H^{(2)}n=0$, the Jacobi field $\xi$ is entirely blind to $\phi_n$, and the normal and tangential dynamics are geometrically decoupled. We now demonstrate that this geometrically decoupled Jacobi field decays exponentially in dissipative systems, severing the macroscopic-microscopic coupling.

\subsubsection{Exponential Cancellation as the Core Mechanism}

The geometric ingredients established in Section~\ref{subsec:contact_constraint} and the Jacobi field equation above set the stage for the central result: the contact constraint forces the effective coupling to vanish. Recall the reduced constraint from \eqref{eq:reduced_constraint}:
\begin{equation}\label{eq:reduced_constraint_restate}
\varepsilon + H^{(0)} = \tfrac{1}{2}\,\phi_\parallel^{\!T} H^{(2)}\,\phi_\parallel = \mathrm{const}.
\end{equation}
Let the eigenvalues of $\tilde{M}$ restricted to $\mathrm{im}\,P_\parallel$ have strictly negative real parts, and define the tangential amplification rate
\begin{equation}\label{eq:tangential_amplification_rate_restate}
\sigma_\parallel := \min_k \bigl(-\mathrm{Re}(\lambda_k^{\tilde{M}})\bigr) > 0 .
\end{equation}
Since the fundamental solution of $\dot{\tilde{\Phi}}=\tilde{M}\tilde{\Phi}$ decays as $e^{-\sigma_\parallel t}$, the tangential block of $H^{(2)}$, and hence $\|H^{(2)}\|$ itself (the normal block vanishing by \eqref{eq:H2_projection_identity}), decays as $e^{-2\sigma_\parallel t}$. The reduced constraint \eqref{eq:reduced_constraint_restate} then forces $|\phi_\parallel|$ to grow as $e^{\sigma_\parallel t}$, and their product vanishes:
\begin{equation}
|H^{(2)}(t)\,\phi_\parallel(t)| \leq \|H^{(2)}(t)\|\,|\phi_\parallel(t)| \sim e^{-2\sigma_\parallel t}\,e^{\sigma_\parallel t} = e^{-\sigma_\parallel t} \to 0 .
\label{eq:effective_perturbation_decay_restate}
\end{equation}
The constraint conservation rigidly couples the growth of $|\phi_\parallel|$ and the decay of $\|H^{(2)}\|$, forcing their product to vanish. The system becomes internally finer ($|\phi_\parallel|\to\infty$, probability field sharpens) while externally blinder ($H^{(2)}\to 0$, coupling softens), so that the macroscopic variable $y$ evolves as if the internal probability field did not exist.

The asymptotic fate of the coupling is governed entirely by the spectrum of $\tilde{M}$. Two primary cases arise. In dissipative systems ($\mathrm{Re}(\lambda_k^{\tilde{M}})<0$), the rate $\sigma_\parallel$ defined in \eqref{eq:tangential_amplification_rate_restate} is strictly positive. The Jacobi field decays as $\xi(t)\sim e^{-\sigma_\parallel t}\to 0$: the dynamical channel linking trajectory to probability field is severed, and determinism emerges via decoherence. This is the dynamical content of the contact constraint. 

In non-dissipative systems ($\mathrm{Re}(\lambda_k^{\tilde{M}})=0$), $\xi$ oscillates coherently without decay; if $\phi(0)=0$ then $\xi\equiv 0$ and the system resides on $\mathcal{S}$ from the outset (a priori determinism), while if $\phi(0)\neq 0$ the trajectory retains permanent coupling to the probability field and deterministic focusing is inapplicable. Having established the core decay mechanism and its spectral conditions, we next examine how this mechanism manifests in different system topologies.

\subsubsection{Two Focusing Regimes and the Deterministic Limit}
The specific manifestation of this exponential cancellation depends on the nature of the zero-order contact potential $H^{(0)}$, leading to two distinct focusing regimes.

When $H^{(0)}=c$ (constant), we have $\nabla H^{(0)}=0$ which gives $P_\parallel=I$, $\phi_\parallel=\phi$, and $\tilde{M}=M$. In this Global Focusing regime, the constraint operates in all directions, forcing the entire probability field to collapse towards a Dirac-like deterministic attractor. The convergence is global, and the base dynamics simplify directly to $\dot{y}^i=H^{(1)i}+H^{(2)ij}\phi_j$.

When $H^{(0)}$ is non-constant, the zero-order transport equation \eqref{eq:zero_order_transport} implies that $H^{(0)}$ is conserved along the characteristics of $H^{(1)}$, so the level sets $\mathcal{M}_c(t):=\{y\mid I(t,y)=c\}$, with $I(t,y):=H^{(0)}(t,y)$, are invariant under the flow. Starting from the original base dynamics equation for the $N=2$ truncation,
\begin{equation}\label{eq:original_base_dynamics}
\dot{y}^i = H^{(1)i}(t,y) + H^{(2)ij}(t,y)\phi_j ,
\end{equation}
we decompose the fibre coordinate $\phi$ using the identity $I = P_\parallel + (I - P_\parallel)$:
\begin{equation}\label{eq:phi_decomposition}
H^{(2)}\phi = H^{(2)}\bigl(P_\parallel\phi + (I - P_\parallel)\phi\bigr) = H^{(2)}(P_\parallel\phi) + H^{(2)}(I - P_\parallel)\phi .
\end{equation}
According to the first-order degeneracy condition \eqref{eq:first_order_degeneracy}, the gradient $\nabla H^{(0)}$ lies in the kernel of $H^{(2)}$, which eliminates the normal projection:
\begin{equation}\label{eq:normal_projection_killed}
H^{(2)}(I - P_\parallel) = H^{(2)} \left(\frac{\nabla H^{(0)} \otimes \nabla H^{(0)}}{|\nabla H^{(0)}|^2}\right) = \frac{1}{|\nabla H^{(0)}|^2} \bigl(H^{(2)}\nabla H^{(0)}\bigr) \otimes \nabla H^{(0)} = \mathbf{0} .
\end{equation}
Substituting \eqref{eq:normal_projection_killed} into \eqref{eq:phi_decomposition} yields $H^{(2)}\phi = H^{(2)}(P_\parallel\phi)$. By the projection identity \eqref{eq:H2_projection_identity}, the base dynamics are strictly confined to the tangent bundle of these level sets:
\begin{equation}\label{eq:tangential_base_dynamics}
\dot{y}^i = H^{(1)i}(t,y) + H^{(2)ij}(t,y)\bigl(P_\parallel\phi\bigr)_j ,
\end{equation}
since $H^{(2)}(P_\parallel\phi)$ is orthogonal to $n$ and $H^{(1)}\cdot\nabla I=-\partial_t I$ from \eqref{eq:zero_order_transport}. This defines the Structured Focusing regime. 

In the autonomous case, $\dot{y}$ is everywhere tangent to $\mathcal{M}_c$, while in the non-autonomous case (e.g. the limit-cycle potential $H^{(0)}(t,y)=\ln\rho_\gamma(y)+\int_0^t\Lambda(\Phi_{-\sigma}(y))\,d\sigma$) the level sets rotate with the drift flow. The convergence operates strictly within $\mathcal{M}_c$, so $\dot{y}=H^{(1)i}(t,y)$ emerges as the geometric attractor within $\mathcal{M}_c$.
Crucially, within this Structured Focusing regime, the normal fibre component $\phi_n$ does not participate in the contact constraint. 

Projecting \eqref{eq:phi_linearized} onto $n$ gives the normal dynamics:
\begin{equation}\label{eq:normal_dynamics}
\frac{d}{dt}(n\cdot\phi) = -|\nabla H^{(0)}| - n^T M^T\phi + \dot{n}\cdot\phi ,
\end{equation}
where the term $n^T H^{(2)} v=0$ for all $v$ (from \eqref{eq:first_order_degeneracy}) ensures that $H^{(2)}$ provides no restoring force in the normal direction. Because $H^{(2)}n=0$, the constraint conservation provides no compensating stiffness for the normal direction, so no contact constraint operates there. The normal component $\phi_n$ is driven by the persistent gradient $-\nabla H^{(0)}$ and does not experience the amplification-decay cancellation; the probability field cannot collapse transversely to the level sets, and the value $c$ of the generalised integral $I(t,y)=H^{(0)}(t,y)$ remains an irreducible uncertainty fixed by the initial condition. 

This reveals a precise geometric trade-off: each independent non-constant $H^{(0)}$ freezes one direction of probability collapse, replacing the global $\delta$-attractor with a singular measure supported on the level set $\mathcal{M}_c$. Determinism is not lost, but geometrically compartmentalised: it governs the drift within the level sets, while the transverse coordinate remains an irreducible parameter of the initial data.
For the $N=2$ truncation, the exact macroscopic equation in both regimes is given by \eqref{eq:tangential_base_dynamics}. Applying the asymptotic bound \eqref{eq:effective_perturbation_decay_restate} yields the convergence
\begin{equation}
\dot{y}^i = H^{(1)i}(t,y) + \mathcal{O}\bigl(e^{-\sigma_\parallel t}\bigr) \to H^{(1)i}(t,y), \quad t\to\infty .
\label{eq:deterministic_emergence}
\end{equation}
Therefore, the equation $\dot{y}^i = H^{(1)i}(t,y)$ is not an approximation but the strict geometric attractor of the contact flow as $t\to\infty$. 

We call this process deterministic focusing, the convergence of the macroscopic trajectory to a deterministic path, driven by the geometric filtering of internal probability structure. The next subsection examines what this convergence means physically.

\subsection{Emergence of Determinism}
\label{subsec:Emergence_determinism}

The convergence $\dot{y}^i \to H^{(1)i}(t,y)$ reveals the physical identity of the first-order contact potential and, with it, the mechanism by which determinism emerges. In classical physics, the macroscopic evolution of a dissipative system is governed by a drift field $f^i(t,y)$, which is typically postulated on the basis of phenomenological or symmetry considerations. 

In the contact geometric framework, no such postulate is needed. The contact potential $H(t,y,\phi)$ is constructed from the invariant measure of the underlying stochastic dynamics, and its first-order coefficient $H^{(1)i}(t,y)$ arises naturally from this geometric construction. On the zero-fibre section $\phi=0$, the contact Hamiltonian equation $\dot{y}^i = \partial H/\partial \phi_i$ reduces to $\dot{y}^i = H^{(1)i}(t,y)$. The contact constraint then forces the effective perturbation $H^{(2)}\phi_\parallel$ to vanish exponentially, so that the full contact flow converges onto this section. The deterministic drift field is therefore identified as
\begin{equation}
f^i(t,y) = H^{(1)i}(t,y),
\end{equation}
not as an assumption, but as a geometric consequence. The deterministic equation $\dot{y}^i = f^i(t,y)$ is not prescribed but discovered as the strict geometric attractor of the contact flow.

Deterministic emergence is thus realised through a two-step mechanism. First, the contact potential, built from the invariant measure of the stochastic dynamics, naturally produces a first-order coefficient $H^{(1)i}$ that plays the role of the classical drift field. Second, the contact constraint enforces the exponential decay of all stochastic corrections; the growth of probability gradients $|\phi_\parallel|$ is exactly counterbalanced by the decay of the second-order contact stiffness $\|H^{(2)}\|$, funnelling the dynamics onto the trajectory governed by this drift field. This emergence is structurally universal: for any initial $\phi(0)\neq 0$, the system undergoes gradient amplification, which is precisely constrained by the contact conservation law, producing a macroscopic trajectory that converges to the deterministic path independent of the microscopic initial conditions.

The asymptotic convergence above is a strict geometric result; its physical interpretation under persistent stochastic driving is examined in Section~\ref{subsec:stochastic_persistence}.

\section{Examples}\label{sec:examples}

We illustrate the invariant-measure construction of Section~\ref{sec:H0_construction} and the deterministic emergence of Section~\ref{sec:Emergence-Determinism} on four representative systems, one for each regime.

\subsection{Point attractor: linear dissipative system}

Consider the two-dimensional linear system
\begin{equation}
\dot y_1 = -\lambda_1 y_1, \qquad \dot y_2 = -\lambda_2 y_2, \qquad \lambda_1,\lambda_2>0.
\end{equation}
The drift field and its divergence are
\begin{equation}
f = (-\lambda_1 y_1,\, -\lambda_2 y_2), \qquad \Lambda = -(\lambda_1+\lambda_2)<0.
\end{equation}
The origin $y_*=(0,0)$ is a hyperbolic sink; every trajectory converges to $y_*$, so the $\omega$-limit set is $\{y_*\}$ and the ergodic invariant measure is $\mu=\delta_{y_*}$. The regularised density $\rho_\varepsilon\to\delta_{y_*}$ gives $S=\ln\rho_\varepsilon\to\mathrm{const}$ in the distributional limit, and $\alpha=-\Lambda=\lambda_1+\lambda_2$. The master formula yields
\begin{equation}
H^{(0)}=c.
\end{equation}
The Jacobian $M=\mathrm{diag}(-\lambda_1,-\lambda_2)$ has spectrum $\{-\lambda_1,-\lambda_2\}$, so $\nabla H^{(0)}=0$ gives $P_\parallel=I$, $\tilde M=M$, and
\begin{equation}
\sigma_\parallel = \min(\lambda_1,\lambda_2)>0.
\end{equation}
This is the global focusing regime: the probability field collapses to $\delta(y-y_*)$ and the macroscopic trajectory converges as $\dot y = f+\mathcal{O}(e^{-\sigma_\parallel t})\to f$.

\subsection{Limit cycle: van der Pol oscillator}

The van der Pol oscillator
\begin{equation}
\ddot x - \mu(1-x^2)\dot x + x = 0, \qquad \mu>0,
\end{equation}
in the phase variables $y_1=x$, $y_2=\dot x$ reads
\begin{equation}
\dot y_1 = y_2, \qquad \dot y_2 = \mu(1-y_1^2)y_2 - y_1.
\end{equation}
The drift field and its divergence are
\begin{equation}
f = \bigl(y_2,\; \mu(1-y_1^2)y_2-y_1\bigr), \qquad \Lambda(y)=\mu(1-y_1^2),
\end{equation}
so the divergence is state-dependent. The system possesses a unique stable limit cycle $\gamma$ of period $T$. The ergodic invariant measure is the equitime measure $\mu_\gamma$ on $\gamma$, with density
\begin{equation}
\rho_\gamma(y) = \frac{1}{T\,|f(y)|}, \qquad y\in\gamma,
\end{equation}
where $|f|=\sqrt{y_2^2+[\mu(1-y_1^2)y_2-y_1]^2}$. The construction gives
\begin{equation}
S(y)=\ln\rho_\gamma(y)=-\ln T-\ln|f(y)|, \qquad \alpha(y)=-\mu(1-y_1^2),
\end{equation}
and the master formula yields
\begin{equation}
H^{(0)}(t,y)=\ln\rho_\gamma(y)+\int_0^t\mu\bigl(1-y_1(\Phi_{-\sigma}(y))^2\bigr)\,d\sigma.
\end{equation}
Since $|f|$ varies along $\gamma$, $\nabla H^{(0)}\neq 0$, and the level sets $\mathcal{M}_c(t)$ are one-dimensional curves rotating with the drift flow. The Jacobian
\begin{equation}
M=\begin{pmatrix}0&1\\-2\mu y_1 y_2-1&\mu(1-y_1^2)\end{pmatrix}
\end{equation}
has, on $\gamma$, one zero Floquet exponent (tangent to $\gamma$) and one transverse exponent
\begin{equation}
\lambda_\parallel = \frac{1}{T}\int_0^T \mu(1-y_1^2)\,dt = \mu\,\overline{(1-y_1^2)}_\gamma < 0,
\end{equation}
negative because the limit cycle has amplitude approximately $2$, so $\overline{y_1^2}_\gamma>1$ and $\overline{(1-y_1^2)}_\gamma<0$. The projected Jacobian $\tilde M=P_\parallel M P_\parallel$ acts on the tangent to $\mathcal{M}_c$; since $\nabla\ln\rho_\gamma=-\nabla\ln|f|$ has a tangential component, the tangent space of $\mathcal{M}_c$ excludes the zero Floquet direction, giving
\begin{equation}
\sigma_\parallel = -\lambda_\parallel > 0.
\end{equation}
This is the structured focusing regime: the probability field collapses to the rotating level set $\mathcal{M}_c(t)$, and within $\mathcal{M}_c$ the trajectory converges as $\dot y = f+\mathcal{O}(e^{-\sigma_\parallel t})\to f$, asymptotically settling onto $\gamma$.

\subsection{Chaos: Lorenz system}

The Lorenz system
\begin{equation}
\dot y_1 = \sigma(y_2-y_1), \quad \dot y_2 = y_1(\rho-y_3)-y_2, \quad \dot y_3 = y_1 y_2-\beta y_3
\end{equation}
with $(\sigma,\rho,\beta)=(10,28,8/3)$ has drift field
\begin{equation}
f = \bigl(\sigma(y_2-y_1),\; y_1(\rho-y_3)-y_2,\; y_1 y_2-\beta y_3\bigr)
\end{equation}
and constant divergence
\begin{equation}
\Lambda = -(\sigma+1+\beta) = -\frac{41}{3}<0.
\end{equation}
The system possesses a strange attractor supporting a unique SRB measure $\mu_{\mathrm{SRB}}$ \cite{Benedicks1993}, selected by Lebesgue-almost-every initial condition. Its density $\rho_{\mathrm{SRB}}=d\mu_{\mathrm{SRB}}/d\Omega$ exists along the unstable directions. The construction gives
\begin{equation}
S(y)=\ln\rho_{\mathrm{SRB}}(y), \qquad \alpha = -\Lambda = \sigma+1+\beta = \frac{41}{3},
\end{equation}
and the master formula yields
\begin{equation}
H^{(0)}(t,y)=\ln\rho_{\mathrm{SRB}}(y)-\frac{41}{3}\,t.
\end{equation}
The SRB density is non-constant, so $\nabla H^{(0)}\neq 0$ and the level sets $\mathcal{M}_c(t)$ are two-dimensional surfaces rotating with the drift flow. The Jacobian
\begin{equation}
M=\begin{pmatrix}-\sigma&\sigma&0\\\rho-y_3&-1&-y_1\\y_2&y_1&-\beta\end{pmatrix}
\end{equation}
has one positive, one negative, and one near-zero Lyapunov exponent. The projected Jacobian $\tilde M=P_\parallel M P_\parallel$ on the tangent to $\mathcal{M}_c$ excludes the direction of $\nabla H^{(0)}$; the negative average divergence $\Lambda<0$ together with the SRB structure ensures that the transverse spectrum has negative real part, giving
\begin{equation}
\sigma_\parallel>0.
\end{equation}
This is again the structured focusing regime: the probability field collapses to $\mathcal{M}_c(t)$, and the trajectory converges as $\dot y = f+\mathcal{O}(e^{-\sigma_\parallel t})\to f$ on the strange attractor.

\subsection{Conservative: harmonic oscillator}

The harmonic oscillator
\begin{equation}
\dot q = p, \qquad \dot p = -\omega^2 q
\end{equation}
in the phase variables $y=(q,p)$ has drift field and divergence
\begin{equation}
f=(p,\,-\omega^2 q), \qquad \Lambda = \frac{\partial p}{\partial q}+\frac{\partial(-\omega^2 q)}{\partial p}=0.
\end{equation}
The trajectories are the ellipses $\omega^2 q^2+p^2=\mathrm{const}$; on each ellipse the ergodic invariant measure is the Liouville measure $\mu_{\mathrm{L}}=dq\,dp$, which coincides with the reference measure $\Omega=dq\,dp$. The construction gives
\begin{equation}
\rho_{\mathrm{L}}=\frac{d\mu_{\mathrm{L}}}{d\Omega}=1, \qquad S=\ln\rho_{\mathrm{L}}=0, \qquad \alpha=-\Lambda=0,
\end{equation}
and the master formula yields the zero-order contact potential
\begin{equation}
H^{(0)}=c.
\end{equation}
The Jacobian
\begin{equation}
M=\begin{pmatrix}0&1\\-\omega^2&0\end{pmatrix}
\end{equation}
has eigenvalues $\lambda=\pm i\omega$, purely imaginary. Hence $\mathrm{Re}(\lambda_k^{\tilde M})=0$ and
\begin{equation}
\sigma_\parallel=0.
\end{equation}
No deterministic focusing occurs. If $\phi(0)=0$ the system resides on the deterministic manifold $\mathcal{S}$ from the outset (a priori determinism); if $\phi(0)\neq 0$ the Jacobi field $u$ oscillates coherently without decay, the trajectory retains permanent coupling to the probability field, and the convergence $\dot y\to f$ does not hold.

\section{Discussions}\label{sec:Discussion}

\subsection{Stochastic Persistence in Real World}\label{subsec:stochastic_persistence}

The deterministic focusing result establishes deterministic focusing as a geometric approach within the truncated contact manifold. However, a critical distinction must be drawn between the mathematical asymptotic limit and physical reality.

Mathematically, deterministic focusing corresponds to the probability density localising into a Dirac $\delta$-distribution, $\rho(y) \to \delta(y - y_{\text{det}})$, with $|\phi| \to \infty$, $H^{(2)} \to 0$, and their product vanishing exactly as $\sim e^{-\sigma_\parallel t}$. In physical reality, however, dissipative systems are perpetually subject to stochastic perturbations, such as thermal fluctuations, environmental coupling, intrinsic microscopic noise, that prevent the probability density from ever collapsing into a true $\delta$-distribution. Instead, the dynamics reaches a non-equilibrium steady state: the fibre coordinate $\phi$ saturates at a large but finite value (its exponential growth balanced by stochastic driving), $H^{(2)}$ decays to a small but finite floor, and the effective coupling $H^{(2)}\phi$ becomes exponentially small but strictly non-zero, manifesting as persistent stochastic jitter on macroscopic trajectories.

The contact equation $\dot{y}^i = f^i(y) + H^{(2)ij}\phi_j$ is, in exact terms, a stochastic differential equation: $f^i(y)$ is the deterministic drift, and $\xi^i=H^{(2)ij}\phi_j$ is the stochastic forcing arising from the transduction of the inhomogeneous probability field into effective dynamic forces. In the contact geometric framework, these disturbances are not appended as an ad hoc noise term; rather, they are encoded geometrically through the fibre coordinate $\phi$, whose dynamics are governed by the conjugate equation $\dot{\phi}_i = -\partial H/\partial y^i$ within the contact system. In the mathematical limit, the contact constraint ensures this perturbation vanishes exponentially; in physical reality, persistent external driving maintains a non-zero $\phi$, and the residual coupling $\xi^i = H^{(2)ij}\phi_j$ is the sole remnant that prevents complete decoupling. The deterministic drift equation $\dot{y} = f(t,y)$ thus provides the exact geometric skeleton and quantifies the gradient amplification timescale $\tau_f = 1/\sigma$; physical systems settle near this attractor within $\mathcal{O}(\tau_f)$, maintaining a finite-width probability field sustained by persistent perturbations.

A crucial dimension of this physical reality is the competition between the focusing timescale $\tau_f = 1/\sigma$ and the characteristic timescale of macroscopic system evolution or external driving, denoted $\tau_{\mathrm{sys}}$. Deterministic focusing is not an instantaneous projection but a dynamic process of geometric reorganisation that requires the system to evolve sufficiently long for the effective coupling to decay. If the system undergoes macroscopic structural changes, traverses rapidly varying regions of phase space, or is subjected to intermittent macroscopic perturbations on a timescale $\tau_{\mathrm{sys}} \lesssim \tau_f$, the focusing process is perpetually interrupted. The system is effectively ``rebooted'' before the macroscopic-microscopic coupling $\xi^i = H^{(2)}\phi$ can vanish. In this regime, the stochastic forcing remains macroscopically relevant indefinitely, leading to a state of long-term stable stochasticity rather than asymptotic determinism. This timescale disparity provides a geometric explanation for why systems near bifurcation points (where eigenvalues approach the imaginary axis, causing $\sigma \to 0$ and $\tau_f \to \infty$) or those subjected to rapid environmental modulation inherently exhibit pronounced, persistent fluctuations. Thus, the observability of macroscopic determinism is fundamentally contingent on whether the system is granted sufficient time, $\tau_f$, to complete the geometric reorganisation of information before the next macroscopic transition occurs.

\subsection{A Mechanical Analogy of the Focusing Mechanism}
\label{subsec:mechanical_analogy}

In Section~\ref{sec:Emergence-Determinism}, the emergence of determinism is established through the decay of the Jacobi field $\xi^i = H^{(2)ij}\phi_j = \dot{y}^i - f^i$. By interpreting this geometric deviation as a macroscopic momentum $p:= \xi = \dot{y} - f$, the contact constraint reveals a Hamiltonian structure that provides a profound mechanical reinterpretation of the focusing mechanism.

For the $N=2$ truncation, the conserved constraint function is
\begin{equation*}
\varepsilon = -H^{(0)}(t,y) + \frac{1}{2} \phi^T H^{(2)}(t,y) \phi.
\end{equation*}
Assuming $H^{(2)}$ is invertible on its image---which strictly holds in the Global Focusing regime $H^{(0)}=c$ where $H^{(2)}$ is full-rank, while in the Structured Focusing regime $\nabla H^{(0)}\neq 0$ the matrix $H^{(2)}$ degenerates in the normal direction ($H^{(2)}n=0$) so that $(H^{(2)})^{-1}$ exists only on the tangential subspace $\mathrm{im}\,P_\parallel$---we substitute $\phi = (H^{(2)})^{-1} p$ to obtain an equivalent representation:
\begin{equation}
\varepsilon = -H^{(0)}(t,y) + \frac{1}{2} p^T \left(H^{(2)}\right)^{-1} p.
\label{eq:hamiltonian_constraint}
\end{equation}
By defining the effective mass tensor $G(t,y) := H^{(2)}(t,y)$, this takes the exact form of a classical energy conservation law:
\begin{equation}
\varepsilon = \underbrace{-H^{(0)}(t,y)}_{\text{Potential Energy}} + \underbrace{\frac{1}{2} p^T G^{-1}(t,y) p}_{\text{Kinetic Energy}}.
\label{eq:energy_conservation}
\end{equation}

This algebraic transformation exposes the underlying mechanics of the stochastic-to-deterministic transition. The zero-order potential $H^{(0)}$ acts as the potential energy of the deterministic background, while the second-order contact stiffness $H^{(2)}$ ceases to be merely a geometric tensor; it provides the inertial mass $G$ that couples the macroscopic trajectory to the underlying probability gradients.

Under this mechanical analogy, the deterministic focusing mechanism acquires a physical interpretation. The asymptotic decay $H^{(2)} \to 0$ (with $|\phi| \to \infty$) corresponds precisely to the dissipation of the effective geometric mass ($G \to 0$). Because the total energy $\varepsilon$ is strictly conserved along the flow, the vanishing of mass forces the geometric momentum $p = G\phi$ to decay exponentially to zero, which leads to $dP/dt=0$ along $y(t)$, implying an equiprobability state. 

Consequently, the system loses its memory of microscopic initial conditions not through statistical averaging or information loss, but because the geometric mass carrying that information has dissipated. This mass dissipation freezes the macroscopic trajectory onto the deterministic manifold $\dot{y} = f$, providing a mechanical foundation for the emergence of certainty.

\subsection{Comparison with the Mori--Zwanzig Projection Method}\label{subsec:MZ_comparison}

 The contact-geometric structure revealed in this paper stands in fundamental contrast to the Mori--Zwanzig projection formalism \cite{Mori1965, Zwanzig1960, Zwanzig1961}. The difference is not merely technical; it is ontological. The two frameworks represent fundamentally different ways of seeing the relationship between stochasticity and determinism.

 We identify three fundamental distinctions: geometry, emergence, and probability, which together reveal the fundamentally different nature of the two frameworks.

\subsubsection{Geometry: Intrinsic Structure vs.\ Operator Projection}

The MZ method begins by partitioning the full state space into ``relevant" (slow, observed) variables and ``irrelevant" (fast, unobserved) variables through a linear projection operator $\mathcal{P}$. This partition is externally imposed: the choice of which variables are relevant is a modelling decision, not a consequence of the dynamics itself. The projection operator $\mathcal{P}$ is typically defined as a conditional expectation with respect to some reference measure, and the complementary projector $\mathcal{Q} = I - \mathcal{P}$ captures the orthogonal (irrelevant) subspace. The geometric content of this procedure is straightforward: it is an algebraic decomposition of the Hilbert space of observables, with no intrinsic connection to the phase-space geometry of the underlying dynamical system.

By contrast, the contact geometric framework does not impose an external partition. The decomposition of the dynamics into a deterministic manifold and stochastic perturbations arises intrinsically from the contact structure of the stochastic vector bundle. The fibre coordinate $\phi_i$ is not a freely chosen projection direction; it is the canonical conjugate variable on the contact manifold $(\mathcal{E}, \Theta)$, determined entirely by the jet-bundle structure of the probability measure space. The zero-fibre section $\phi = 0$ is a geometrically distinguished submanifold, i.e., the deterministic manifold, not an arbitrary slicing of the state space. The contact 1-form $\Theta = H\,dt - \phi_i\,dy^i$ and its non-closure $d\Theta \neq 0$ encode the irreversibility and dissipation of the system as geometric invariants. At the same time, the constraint function $\varepsilon = D[H]$ provides a conserved quantity that governs the entire dynamics through the least constraint variational principle. In short, the contact framework replaces the externally imposed projection $\mathcal{P}$ of MZ with an intrinsically geometric decomposition dictated by the contact manifold itself.

\subsubsection{Emergence: Geometric Attractor vs.\ Statistical Approximation}

The MZ procedure yields a generalised Langevin equation for the relevant variables:
\begin{equation}
\frac{dA(t)}{dt} = \Omega\, A(t) + \int_0^t K(t-s)\,A(s)\,ds + F(t), \label{eq:MZ_GLE}
\end{equation}
where $\Omega$ is the Markovian (frequency) matrix, $K(t-s)$ is the memory kernel encoding the delayed influence of the irrelevant variables, and $F(t)$ is the orthogonal dynamics (stochastic forcing). The deterministic averaged equation is recovered only by discarding the memory integral and the stochastic forcing, that is, by the Markovian closure approximation $K \equiv 0$, $F \equiv 0$. This closure is not exact; it is a statistical approximation whose validity depends on the separation of timescales between relevant and irrelevant variables. When timescale separation fails, the memory kernel cannot be neglected, and the deterministic equation is at best an approximation.

In the contact geometric framework, deterministic focusing is not an approximation but a strict geometric result. The exact macroscopic equation $\dot{y}^i = f^i(t,y) + H^{(2)ij}(t,y)(P_\parallel\phi)_j$ converges to $\dot{y}^i = f^i(t,y)$ because the effective coupling $H^{(2)}\phi_\parallel$ vanishes exponentially as a consequence of the contact constraint mechanism. No closure approximation is involved; the conservation of the constraint function $\varepsilon$ guarantees that the gradient amplification ($|\phi_\parallel| \sim e^{\sigma_\parallel t}$) is exactly counterbalanced by the stiffness decay ($\|H^{(2)}\| \sim e^{-2\sigma_\parallel t}$), forcing the back-reaction to vanish as $e^{-\sigma_\parallel t}$. The deterministic equation $\dot{y} = f(t,y)$ is the geometric attractor of the contact flow, not a truncated approximation of a more general equation. Furthermore, the MZ framework requires the user to pre-select the relevant variables; determinism is assumed rather than derived. In the contact framework, the deterministic manifold is identified as the asymptotic attractor; the system reveals which dynamics emerge, rather than the modeller prescribing them.

\subsubsection{Probability: Conservation Law vs.\ Fluctuation--Dissipation}

The role of probability in the two frameworks is fundamentally different. In the MZ formalism, probability enters through the initial distribution of the irrelevant variables and the fluctuation--dissipation relation that connects the memory kernel $K(t)$ to the correlation of the stochastic forcing $F(t)$. The probability field is external to the dynamical equations; it is a statistical input required to close the system. The MZ equation itself is an identity (exact at the level of the full Liouville equation), but its practical utility depends entirely on statistical assumptions about the irrelevant subspace; typically, that the irrelevant variables are initially Gaussian and rapidly decorrelating.

In the contact geometric framework, probability is not an external statistical input but an intrinsic geometric degree of freedom. The fibre coordinate $\phi_i$ is constructed as a weighted sum of all orders of spatial derivatives of the probability distribution $P$, so the probability field is woven into the fabric of the contact manifold itself. The conservation of the constraint function $\varepsilon$ is not a statistical assumption but a geometric conservation law; it follows from the contact Hamiltonian flow preserving the contact structure ($\mathcal{L}_{X_H}\Theta = 0$). The contact constraint mechanism, which is the engine of deterministic focusing, is a direct consequence of this conservation law: the constraint $\tfrac{1}{2}\phi_\parallel^T H^{(2)}\phi_\parallel = \varepsilon + H^{(0)}$ forces the growth of $|\phi_\parallel|$ and the decay of $\|H^{(2)}\|$ to be constrained in exact opposition. No assumption of Gaussianity, rapid decorrelation, or timescale separation is required. The probability field is not averaged away or projected out; it is geometrically decoupled by the contact dynamics itself: the system becomes internally finer while becoming blinder externally.

 The following table consolidates these distinctions:

\begin{table}[htbp]
\centering
\renewcommand{\arraystretch}{1.4} 
\small                          
{\color{blue}\caption{Mori--Zwanzig Projection vs.\ Contact-Geometric Structure: Two Ways of Observing}
\label{tab:MZ_vs_Contact}}
\begin{tabularx}{\textwidth}{@{} >{\raggedright\arraybackslash}p{3.5cm} X X @{}}
\toprule
\textbf{Aspect} & \textbf{Mori--Zwanzig} & \textbf{Contact Geometry} \\ 
\midrule
Decomposition 
  & Externally imposed projection $\mathcal{P}$ 
  & Intrinsic contact structure $\Theta$ \\
\addlinespace 

Relevant variables 
  & Pre-selected by the modeller 
  & Discovered as a geometric attractor \\
\addlinespace 

Deterministic equation 
  & Markovian closure approximation 
  & Strict geometric consequence \\
\addlinespace 

Memory effects 
  & Memory kernel $K(t-s)$, requires closure 
  & Embedded in $\phi$ via cumulant-based coefficients and higher-order probability derivatives \\
\addlinespace 

Probability 
  & External statistical input (initial distribution, FDT) 
  & Intrinsic geometric degree of freedom ($\phi_i$) \\
\addlinespace 

Emergence mechanism 
  & Timescale separation + statistical averaging 
  & Constraint conservation + geometric reorganisation \\
\addlinespace 

Validity 
  & Requires separation of timescales 
  & Requires dissipative spectrum ($\mathrm{Re}(\lambda_k) < 0$) \\
\bottomrule
\end{tabularx}
\end{table}

\section{Conclusions}\label{sec:conclusions}

This paper analyzes how deterministic macroscopic dynamics arises from stochastic fluctuations in dissipative systems using contact geometry, and the main results are summarised as follows:

\begin{enumerate}[label=(\roman*)]

    \item For dissipative systems, deterministic trajectories form geometric attractors of contact flows. The contact constraint balances the exponential growth of probability gradients with the exponential decay of second-order contact stiffness, leading to exponentially weak coupling between macroscopic motion and microscopic fluctuations. This process does not require separation of time scales or Gaussian assumptions about noise.
    
    \item The coupling term linking system states and probability gradients is a Jacobi field of the drift flow. This field decays exponentially in dissipative systems and allows deterministic focusing. In conservative systems, the Jacobi field oscillates without decay, so full separation from stochastic fluctuations cannot be realised.

    \item This contact geometric framework differs from the Mori–Zwanzig projection method. Mori–Zwanzig filters out irrelevant variables and approximates memory terms, which discard information. The contact geometric approach keeps the constraint function conserved, and deterministic behaviour comes from geometric rearrangement of stochastic dynamics instead of information loss.
    
    \item The invariant-measure construction provides a unified way to construct zero-order contact potentials, linking local volume contraction to the global topological features of attractors. This framework theoretically verifies the above mechanism for point attractors, limit cycles, and strange attractors.

\end{enumerate}

\section*{List of Main Notations}
\label{sec:notations}

\noindent \textbf{Latin Letters} \\[0.5em]
$D$: Discriminant operator ($E - 1$) \\
$E$: Total space of the stochastic vector bundle \\
$E$: Euler operator ($\phi_i \partial / \partial \phi_i$) \\
$\{F, G\}$: Canonical Poisson bracket \\
$f^i(t,y)$: First-order contact potential coefficient ($f^i = H^{(1)i}$), identified as the drift field ($\dot{y}^i|_{\phi=0}$, Section~\ref{subsec:deterministic_focusing}) \\
$H$: Contact probability potential (rate of change of the probability measure along evolution trajectories) \\
$H^{(0)}$: Zero-order contact potential (constant $c$ when $\nabla H^{(0)} = 0$; generalized integral $I(t,y)$ when $\nabla H^{(0)} \neq 0$) \\
$H^{(1)}$: First-order contact potential (first-order term $H^{(1)i} \phi_i$) \\
$H^{(2)}$: Second-order contact potential (second-order contact stiffness tensor) \\
$H^{(n)}$: $n$-th order homogeneous polynomial term in the Taylor expansion of $H$ with respect to $\phi$ \\
$I(t,y)$: Generalized integral (equivalent to $H^{(0)}(t,y)$) \\
$J^\infty(E, \mathbb{P})$: Infinite-order jet bundle of the probability measure space \\
$\ker(d\Theta)$: Subspace of vector fields preserving the contact structure \\
$\ker(\Theta)$: Contact distribution (probability evolution space) \\
$M$: $m$-dimensional smooth base manifold (physical state space) \\
$M^i_j(t,y)$: Drift-field Jacobian matrix ($\partial f^i / \partial y^j$) \\
$\widetilde{M}$: Projected drift-field Jacobian ($P_\parallel M P_\parallel$) \\
$n$: Unit normal vector to the level sets of $H^{(0)}$ ($\nabla H^{(0)}/|\nabla H^{(0)}|$) \\
$(\Omega, \mathcal{F}, \mathbb{P})$: Complete filtered probability space \\
$P_\parallel$: Orthogonal projector onto the tangent space of $H^{(0)}$ level sets \\
$\mathcal{S}$: Zero-fiber section ($\phi=0$, deterministic manifold of the system) \\
$S$: Constraint action functional ($\int \varepsilon \, dt$) or geometric potential ($\ln(d\mu/d\Omega)$, context-dependent) \\
$t$: Evolution parameter (time variable) \\
$\xi^i(t, y, \phi)$: Jacobi field of the drift flow ($H^{(2)ij}\phi_j$) \\
$X_H$: Contact vector field (unique smooth vector field preserving the contact structure) \\
$y^i$: Base coordinates (state variables of the stochastic system) \\

\vspace{1em}
\noindent \textbf{Greek Letters} \\[0.5em]
$\alpha(y)$: Source term in the transport equation ($\alpha = -\mathcal{L} \ln \Omega$) \\
$\Delta[H^{(2)}]$ \text{or} $\Delta$: Rotational compensator (tracks rotation of the constraint direction $\nabla H^{(0)}$) \\
$\varepsilon$: Constraint function ($D[H] = \phi_i \dot{y}^i - H$, geometric invariant encoding dissipation and stochastic noise) \\
$\lambda_k$: Eigenvalues of the drift-field Jacobian matrix $M$ \\
$\Lambda(t,y)$: Divergence of the drift field ($\nabla \cdot f$; negative on average for dissipative systems) \\
$\phi_i$: Fibre coordinates (conjugate variables encoding stochastic fluctuation and dissipation) \\
$\pi: E \to M$: Bundle projection map \\
$\sigma$: Gradient amplification rate ($\max_k \mathrm{Re}(-\lambda_k) > 0$ for dissipative systems) \\
$\sigma_\parallel$: Tangential amplification rate restricted to the level sets of $H^{(0)}$ \\
$\tau_f$: Characteristic timescale for gradient amplification ($1/\sigma$) \\
$\Theta$: Canonical contact 1-form ($H \, dt - \phi_i \, dy^i$) \\
$\widetilde{\Phi}$: Fundamental solution matrix of the projected variational equation \\
$\Omega$: Canonical volume form induced by coordinates $y^i$ \\

\bibliography{bib-Contact-2026.bib} 

@article{Benedicks1993,
	author = {Benedicks, Michael and Young, Lai-Sang},
	date = {1993/12/01},
	doi = {10.1007/BF01232446},
	id = {Benedicks1993},
	isbn = {1432-1297},
	journal = {Inventiones mathematicae},
	number = {1},
	pages = {541--576},
	title = {Sinai-Bowen-Ruelle measures for certain H{\'e}non maps},
	url = {https://doi.org/10.1007/BF01232446},
	volume = {112},
	year = {1993}}

@article{Erdmann2000Brownian,
	author = {Erdmann, U. and Ebeling, W. and Schimansky-Geier, L. and Schweitzer, F.},
	doi = {10.1007/s100510051104},
	journal = {The European Physical Journal B},
	number = {1},
	pages = {105--113},
	title = {Brownian particles far from equilibrium},
	volume = {15},
	year = {2000}}

@book{nicolis1977Self-organization,
	address = {New York},
	author = {Nicolis, G. and Prigogine, I.},
	isbn = {0-471-02401-5},
	publisher = {John Wiley \& Sons},
	title = {Self-Organization in Nonequilibrium Systems: From Dissipative Structures to Order through Fluctuations},
	year = {1977}}

@article{Perc2006From,
	author = {Perc, M. and Gosak, M. and Marhl, M.},
	doi = {10.1016/j.cplett.2006.01.065},
	journal = {Chemical Physics Letters},
	number = {1--3},
	pages = {106--110},
	title = {From stochasticity to determinism in the collective dynamics of diffusively coupled cells},
	volume = {421},
	year = {2006}}

@article{Zwanzig1961,
	author = {Zwanzig, R.},
	doi = {10.1103/PhysRev.124.983},
	journal = {Physical Review},
	number = {4},
	pages = {983--992},
	title = {Memory effects in irreversible thermodynamics},
	volume = {124},
	year = {1961}}

@article{Zwanzig1960,
	author = {Zwanzig, R.},
	doi = {10.1063/1.1731409},
	journal = {The Journal of Chemical Physics},
	number = {5},
	pages = {1338--1341},
	title = {Ensemble method in the theory of irreversibility},
	volume = {33},
	year = {1960}}

@article{Mori1965,
	author = {Mori, H.},
	doi = {10.1143/PTP.33.423},
	journal = {Progress of Theoretical Physics},
	number = {3},
	pages = {423--455},
	title = {Transport, collective motion, and {Brownian} motion},
	volume = {33},
	year = {1965}}

@article{Cross1993,
	author = {Cross, M. C. and Hohenberg, P. C.},
	doi = {10.1103/RevModPhys.65.851},
	journal = {Reviews of Modern Physics},
	number = {3},
	pages = {851--1112},
	title = {Pattern formation outside of equilibrium},
	volume = {65},
	year = {1993}}

@article{ZHONG2025,
	author = {Zhong, D. Y. and Wang, G. Q.},
	doi = {10.1016/j.chaos.2025.117143},
	journal = {Chaos, Solitons \& Fractals},
	pages = {117143},
	title = {Least constraint and contact dynamics of stochastic vector bundles},
	volume = {201},
	year = {2025}}

@book{anold2010,
	address = {New York},
	author = {Arnold, V. I.},
	doi = {10.1007/978-1-4757-2063-1},
	edition = {2},
	publisher = {Springer},
	series = {Graduate Texts in Mathematics},
	title = {Mathematical Methods of Classical Mechanics},
	volume = {60},
	year = {1989}}

@book{Pope_2000,
	address = {Cambridge},
	author = {Pope, S. B.},
	doi = {10.1017/CBO9780511840531},
	publisher = {Cambridge University Press},
	title = {Turbulent Flows},
	year = {2000}}

@article{Zhong_2024,
	author = {Zhong, D. Y. and Wang, G. Q.},
	doi = {10.1088/1751-8121/ad483a},
	journal = {Journal of Physics A: Mathematical and Theoretical},
	number = {22},
	pages = {225004},
	title = {Kinetic equation for stochastic vector bundles},
	volume = {57},
	year = {2024}}

@article{ZHONG-2022-KINETICEQAUTION,
	author = {Zhong, D. Y. and Wang, G. Q. and Zhang, M. X. and Li, T. J.},
	doi = {10.1063/5.0011056},
	journal = {Physics of Fluids},
	number = {7},
	pages = {073301},
	title = {Kinetic equation for particle transport in turbulent flows},
	volume = {32},
	year = {2020}}

@book{ArnoldVI1989,
	address = {Geneva},
	author = {Arnold, V. I.},
	number = {34},
	publisher = {L'Enseignement Math{\'e}matique},
	series = {Monographies de L'Enseignement Math{\'e}matique},
	title = {Contact Geometry and Wave Propagation: Lectures Given at the University of Oxford under the Sponsorship of the International Mathematical Union},
	year = {1989}}

@book{van1992stochastic,
	address = {Amsterdam},
	author = {van Kampen, N. G.},
	edition = {2},
	isbn = {0-444-89349-0},
	publisher = {North-Holland},
	title = {Stochastic Processes in Physics and Chemistry},
	year = {1992}}

@book{zwanzig2001nonequilibrium,
	address = {New York},
	author = {Zwanzig, R.},
	doi = {10.1093/oso/9780195140187.001.0001},
	publisher = {Oxford University Press},
	title = {Nonequilibrium Statistical Mechanics},
	year = {2001}}

@book{Abraham1978,
	address = {Reading, Massachusetts},
	author = {Abraham, R. and Marsden, J. E.},
	edition = {2},
	isbn = {0-8053-0102-X},
	publisher = {Benjamin/Cummings},
	title = {Foundations of Mechanics},
	year = {1978}}

@article{Bravetti2017,
	author = {Bravetti, A. and Cruz, H. and Tapias, D.},
	doi = {10.1016/j.aop.2016.11.003},
	journal = {Annals of Physics},
	pages = {17--39},
	title = {Contact {Hamiltonian} mechanics},
	volume = {376},
	year = {2017}}

\end{document}